%% file: UE18_ver_7.tex
\documentclass[aps,prx,twocolumn,showpacs,superscriptaddress,groupedaddress]{revtex4-1}
\usepackage{pgf}
\usepackage{tikz}
\usetikzlibrary{calc,patterns,decorations.pathmorphing,decorations.markings}
\usetikzlibrary{arrows}
\usepackage[caption=false]{subfig}
\usepackage{graphicx} 
\usepackage{mathrsfs}
\usepackage{amsfonts}
\usepackage{amsmath}
\usepackage[noabbrev]{cleveref}
\usepackage[export]{adjustbox}
\usepackage{import}
\usepackage[nice]{units}
\usepackage{color}

  \def\a{\alpha} \def\b{\beta}
 \def\d{\delta} \def\l{\lambda}

\def\e{\varepsilon}

\def\gbar{\bar{g}}

\def\gbaro{\bar{g}^0}

\def\nn{\nonumber \\}

\begin{document}
\title{The metric description of viscoelasticity and instabilities in viscoelastic solids}
\date{\today}
\author{Erez Y. Urbach}
\author{Efi Efrati} 
\email{efi.efrati@weizmann.ac.il} 
\affiliation{Department of
Physics of Complex Systems, Weizmann Institute of Science, Rehovot
76100, Israel}

\begin{abstract} 
	Many manmade and naturally occurring materials form viscoelastic solids. The increasing use of biologically inspired elastomeric material and the extreme mechanical response these elastomers offer drive the need for a better, more intuitive and quantitative understanding of the mechanical response of such continua. This is in particular important in determining the stability of viscoelastic structure over time where in lieu of robust rules we often must resort to simulations.
	
	In this work we put forward a metric description of viscoelasticity in which the continua is characterized by temporally evolving reference lengths quantified by a rest reference metric. This rest reference metric serves as a state variable describing the result of the viscoelastic flow in the system, and allows us to provide robust claims regarding stability of incompressible isotropic viscoelastic media. We demonstrate these claims for a simple bistable systems of three standard-linear-solid spring dashpot assemblies where the predicted rules can be verified by explicit calculations, and also show quantitative agreement with recent experiments in viscoelastic silicone rubber shells that display delayed stability loss.
	\end{abstract}
\maketitle


\section{Introduction}
Elastomeric solids are ubiquitous materials composed of cross-linked long molecular chains and display extreme mechanical response properties. Examples include latex rubber, silicone rubbers \cite{LT75} as well as the crosslinked proteins resilin and elastin responsible for the mechanical properties of tissues such as ligaments and arteries~\cite{SA63,NM50,GLCG02}.
Elastomeric solids are characterized by low elastic moduli, and reversible deformations even at strains exceeding one hundred percent. However, they also show stress relaxation when held at constant displacement and creep under a constant load~\cite{RC12}. These creep and stress relaxation phenomena are dissipative yet are reversible. Unlike viscoelastic fluids, elastomeric solids retain the topology of material elements in the body indefinitely. Moreover, they display only fractional stress relaxation, supporting a finite fraction of the stress even after arbitrarily long relaxation times. 	

One of the most fascinating phenomena displayed by elastomeric materials is delayed stability loss, in which a fast instability releasing the elastically stored energy in a system is preceded by a slow creep. Such phenomena have been observed in the rapid snapping of the Venus fly-trap leaf ~\cite{FSDM05} activated by the plant to capture prey, in the passive snap through of thin elastomeric shells known as jumping poppers ~\cite{PMVH14,UE18a}, and for non-elastomeric materials in the slow crustal dynamics leading to some earthquakes  \cite{FL01,RSA94}.

The equations of state for viscoelastic materials commonly relate the stress to the full history of the strain in the body through a material dependent memory kernel~\cite{RC12}. Such equations of state accurately capture the material response, and may be used to numerically study the response of viscoelastic structures given their geometry and loading conditions~\cite{MS10,BSP12}. However, they rarely allow explicit solutions and provide very little insight to the state of the viscoelastic material and its general response properties. The lack of intuition for viscoelastic dynamics is exceptionally evident when considering viscoelastic instabilities; as linear stress relaxation acts to reduce local stresses it is expected to have a stabilizing effect and in particular never cause a meta-stable state to lose stability. More elaborated variations of this hand waving argument, as we show in section \ref{sec:stationary_states}, only prove it stronger rather than refute it, thus elucidating the subtle nature of viscoelastic instabilities.  

	In this works we describe a metric approach to viscoelasticity and viscoelastic stability. We begin by describing  one dimensional viscoelastic systems through temporally evolving rest lengths and specifically formulate the dynamics of the standard linear solid (SLS) model as a spring with temporally evolving rest length. We then consider an assembly of three SLS springs in the form of a Von-Misses truss as a first example of transient elastic stability in viscoelastic systems. We conclude the section describing this one dimensional motion by using the temporally evolving rest lengths to prove general claims regarding the elastic stability and stationarity of a general 1D, SLS systems. 
	
	The intuitive results we obtain for one dimensional SLS systems can be generalized to full three dimensional systems. We next do so by constructing a covariant metric description of viscoelastic solids. The theory describes the material response as elastic with respect to a time dependent three dimensional reference metric. We elucidate the notions of quasi static approximation, transient elastic stability and isotropicity in light of this new description. We then discuss results regarding the quasi-static dynamics when considering isotropic and incompressible materials. Several claims regarding the elastic stability and stationary states of such bodies are shown. 

\section{Viscoelasticity and delayed stability loss in 1D}
	\label{sec:1d_viscoelasticity}

	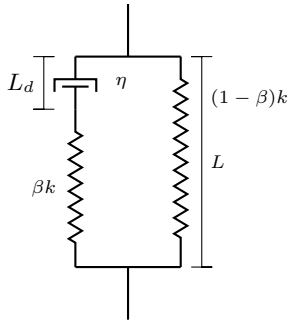
\begin{figure}[ht]
		\scalebox{1}{\import{.}{SLS_diagram.tex}}
		\caption{Standard linear solid with total stiffness $k$, $\b$ and viscosity $\eta$.}
		\label{fig:sls_diagram}
	\end{figure}
	\subsection{Standard linear solid}

		The simplest intuitive model capable of displaying both finite stress relaxation and a finite response at high rates is called the standard linear solid model (SLS) \cite{Zen48}. 
		The SLS is a generalization of two spring-dashpot-models - the Kelvin-Voigt model, accounting for the creep response, and the Maxwell model capturing the stress relaxation and high loading rate response \cite{AB09}. The SLS consist of spring of stiffness $\beta k$ and dashpot of viscosity $\eta$ connected in series, both connected in parallel to another spring of stiffness $(1-\beta)k$ (see FIG.~\ref{fig:sls_diagram}). The dimensionless constant, $0\le \b \le 1$, accounts for difference between the spring stiffness. The total length of the system is denoted $L$, the length on the dashpot is $L_d$, and the rest length of the long and short springs read $\bar{L}^0$ and $\tilde{L}^0$, respectively. The total force is then
		\begin{equation}
			F = -(1-\b)k (L- \bar L ^0) - \b k (L-L_d- \tilde L ^0), \label{eq:sls_force_relation}
		\end{equation}
		while force balance between the spring and dashpot in series yields  the closure relation
		\begin{equation}
		\eta \dot L_d = \b k (L-L_d- \tilde L ^0). \label{eq:sls_closure_relation}
		\end{equation}
		Rapid variation of the total length $L\to L+\Delta L$ will elongate the springs while leaving the dashpot length, $L_{d}$ unchanged. The force under such conditions will increase by $\Delta F = k \Delta L$, similar to a simple spring of stiffness $k$. Motivated by the temporal scale separation between the elastic response of viscoelastic solids and their typical creep rate we seek to rephrase the  force equations as a simple elastic spring. By setting the \emph{reference length} 
		\[
		\bar L = (1-\beta)\bar{L}^0 +\beta\tilde L ^{0}+ \beta L_d,
		\]
		and substituting in eq.\eqref{eq:sls_force_relation} we obtain 
		\begin{equation}
			F = -k (L-\bar L). \label{eq:force_relation}
		\end{equation}
		This is supplemented by the closure relation describing the temporal evolution of the reference length
		\begin{equation}
			\dot {\bar L} = - \frac{1}{\tau}\left( \b (\bar L - L) + (1-\b)(\bar L - \bar L ^0) \right), \label{eq:sls_L_bar_DE}
		\end{equation}
		where $\tau = \frac{\eta}{\b k}$.  The reference length evolves simultaneously towards $L$, the present state of the system, and towards $\bar L^{0}$, the rest length to which it will asymptotically approach if left unconstrained. The two simultaneous evolution terms are weighted by the dimensionless factor $0\le\beta\le 1$. $\b =0$ corresponds to the elastic case in which $\bar L$ starts and stays at $\bar L ^0$. $\b = 1$ corresponds to a Maxwell material where the material reference length $\bar L$ has no preferred rest value and approaches $L$, relaxing the force it supports to zero. One can easily generalize the SLS model to account for multiple relaxation time scales by adding more spring dashpot pairs to the system in parallel, and even account for a distribution of time scales that could give rise to non-exponential relaxation; e.g. power-law or logarithmic. Such generalizations will not change the notion of the reference length or its interpretation through eq.\eqref{eq:force_relation}, but only change the closure relation describing its temporal evolution:
		\begin{equation}
			\bar{L} (t) = (1-\b)\bar L ^0 - \b \int_{-\infty}^{t} \dot \phi(t-s)L (s) ds. \label{eq:sls_lbar_evolution_rule}
		\end{equation}
		Note that $\b$ retains its meaning as the fraction of force that is relaxed asymptotically in a constant displacement setting. As we will see later this quantity dominates most of the questions of stability in viscoelasitc systems, while the actual functional form of the memory kernel, $\phi$, only influences the temporal approach to instability. For this reason it suffices to examine the question of stability loss in 1D considering only the behavior of a simple SLS model. While \cref{eq:force_relation,eq:sls_lbar_evolution_rule} are a mere reformulation of the familiar viscoelastic dynamics, stating the problem in these new variables allows an intuitive interpretation of the dynamics and deeper insight to the behavior such systems exhibit. 

	\begin{figure}[t]
		\scalebox{1}{\import{.}{SLS_model_diagram.tex}}
		\caption{The SLS von-Mises truss model. Point mass connected to two diagonal SLS with $k_1$ each, and a vertical SLS with total $k_2$. Both with dashpot $\eta$ and compliance $\b$.
		}
		\label{fig:sls_von_mises_truss_diagram}
	\end{figure}
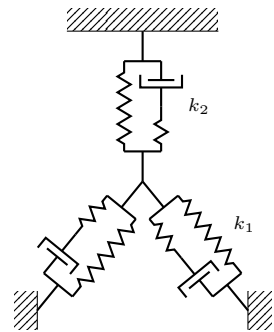

	\subsection{SLS von-Mises truss}
		\label{sec:sls_von_Mises_truss}
\begin{figure}[t]
\resizebox{.9\linewidth}{!}{\includegraphics{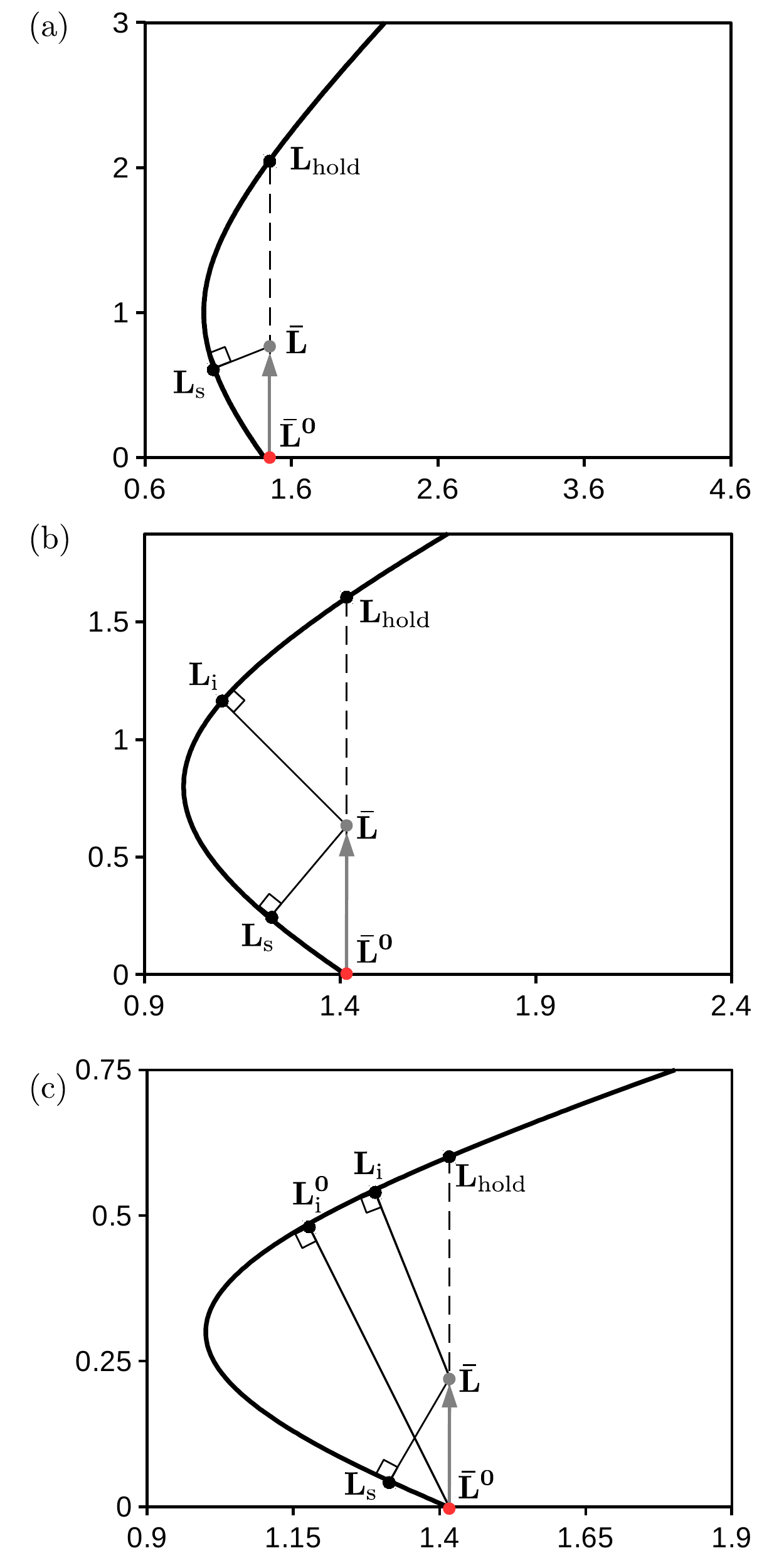}}			
\caption{The normalized lengths 2D space for (a) $\a = 1$, (b) $\a = 0.3$ and (c) $\a = 0.1$. The curve of admissible lengths $\mathbf{L}(z)$ is drawn in thick (black) line. The initial lengths 
			$\mathbf{\bar L}^0=(\sqrt{2},0)$ 
			is drawn in (red) filled circle. During the relaxation the reference lengths $\mathbf{\bar L}$ - drawn in gray filled circle - evolve in the direction of the holding lengths $\mathbf{L}_\text{hold}=\mathbf{L}(z=-1)=(\sqrt{2},2\a)$. The stable positions of a specific reference lengths is geometrically the normals from its point to the curve. (a) Unstable - even along the relaxation the system remain mono-stable. (b) Finitely stable - bistability emerge because of the relaxation, and dissolve finite time after releasing the system. (c) Stable - the system remains bistable during the holding.}
			\label{fig:sls_von_mises_truss_lengths_space}
		\end{figure}

		One of the most striking and subtle viscoelastic phenomena is delayed instability. To capture the essence of the phenomena we consider three SLS's of similar material parameters $\b$ and $\tau$, arranged to form a von-Mises truss \cite{ZPB91} as illustrated in figure~\ref{fig:sls_von_mises_truss_diagram}. 
		The truss is composed of a point mass that moves along the $z$ axis. It is connected to two diagonal SLS's of stiffness $k_1$ that create a bi-stable elastic potential. The symmetry of this potential is broken by a vertical SLS of stiffness $k_2$. The rest lengths are chosen such that the system is relaxed and stationary at position $z = 1$ at rest lengths $\bar L^0_1 = \sqrt{2}$ and $\bar L^0_2 = 0$. The lengths of each spring as function of $z$ are $L_1(z) = \sqrt{1+z^2}$ and $L_2(z) = |z-1|$ respectively. As displayed on the previous section, each SLS is associated with a reference length $\bar L_i$ that slowly evolves according to Eq.~\eqref{eq:sls_L_bar_DE}. The force on the mass is minus the derivative with respect to $z$ of the instanteneous elastic energy
		\begin{equation}
			E(z) =2 \frac{k_1}{2}(L_1(z)-\bar L_1)^2 + \frac{k_2}{2}(L_2(z)-\bar L_2)^2 \label{eq:vmt_energy}
		\end{equation}
		We set $\a = \sqrt{\tfrac{k_2}{2 k_1}}$ and note that for the purely elastic truss bi-stability is obtained for small values of $\a$ whereas mono-stability is displayed at higher $\a$, where only the $z>0$ position is stable.
		We wish to investigate the dynamics of the system in the quasi-static approximation, in which the the mass position is causes vanishing of the total force $\frac{dE}{dz}=0$. For these purposes it is instructive to employ a normalized energy function 
		\begin{equation}
			e(z) = \frac{E(z)}{k_1}  = (L_1(z)-\bar L_1)^2 + (\a L_2(z)-\a \bar L_2)^2 \label{eq:vmt_norm_energy}
		\end{equation}
		This energy function may be interpreted as the Euclidean ``distance'' between the normalized reference lengths $\mathbf{\bar L} = (\bar L_1, \a \bar L_2)$ and the normalized configuration lengths $\mathbf{L}(z) = (L_1(z),\a L_2(z))$. 
		We can thus use this 2D 'phase space' of normalized lengths to understand the behavior of the system. The admissible (normalized) lengths of the springs $\mathbf{L}(z)$ form a one dimensional curve parameterized by $z$, whereas the reference lengths $\mathbf{\bar L}$ could be anywhere in the two dimensional space. The Euclidean distance between the configuration $\mathbf{L}$ and the reference lengths $\mathbf{\bar L}$ is exactly the elastic energy of the configuration.
		Specifically, as $\mathbf{L}(z) = (\sqrt{1+z^2}, \a |z-1|)$, $z<1$ parametrize a hyperbola with its vertex at $(1,\a)$ (corresponding to $z=0$) and asymptotic line $(L_1, \a L_1)$ as $z\rightarrow -\infty$. The rest lengths $\mathbf{\bar L^0}=(\sqrt{2},0)$ is a point on the admissible lengths with $z=1$, or $\mathbf{\bar L^0} = \mathbf{L}(z=1)$ (see FIG.~\ref{fig:sls_von_mises_truss_lengths_space}). 

		To understand the dynamics of the system, we choose the following protocol; the mass is taken from rest, where $L_i=\bar L_i = \bar L_i^0$, abruptly to $z=-1$ and held there for a finite time. That is, at $t=0$ the reference lengths $\mathbf{\bar L}$ are at the rest length values $\mathbf{\bar L^0}=(\sqrt{2},0)$ whereas the springs are held at the configuration $\mathbf{L}_\text{hold} = (\sqrt{2},2 \a)$. By Eq.~\ref{eq:sls_L_bar_DE} during the holding the reference lengths $\mathbf{\bar L}$ evolve along the line connecting $\mathbf{\bar L^0}$ and $\mathbf{L}_\text{hold}$ up to a stationary position between them (See FIG.~\ref{fig:sls_von_mises_truss_lengths_space}). We note that due to the relaxation the reference lengths assume values not on the curve of admissible lengths, $\mathbf{L}(z)$, and therefore do not correspond to any realizable configuration of the system. Nevertheless $\mathbf{\bar L}$ might have stable realizable configurations. An elastic equilibrium corresponds to a point of minimal  ``distance'' according to \eqref{eq:vmt_norm_energy} between $L(z)$ and  $\mathbf{\bar L}(t)$.  Geometrically, this minimization condition is satisfied if the line between the reference lengths $\mathbf{\bar L}$ and the configuration $\mathbf{L}$ is normal to the admissible lengths curve $\mathbf{\bar L(z)}$ (see FIG.~\ref{fig:sls_von_mises_truss_lengths_space}).
		After the release of the mass from $\mathbf{L}_\text{hold}$ it will elastically snap to the closest stable point. Using this geometrical interpretation of the viscoelastic quasi-static evolution, we can explain how different values of $\a$ create different dynamical phases.

We first consider cases with large $\a \rightarrow \infty$, illustrated in FIG.~\ref{fig:sls_von_mises_truss_lengths_space}a. At rest, $\mathbf{\bar L}=\mathbf{\bar L^0}$, only the rest lengths's point $\mathbf{\bar L^0}$ is stable. That is, no other line connecting the point $\mathbf{\bar L^0}$ with the curve $\mathbf{L}(z)$ meets the curve $\mathbf{L}(z)$ perpendicularly. During the holding, the reference lengths $\mathbf{\bar L}$ approache $\mathbf{L}_\text{hold}$. However, still there is only one stable distance minimizing solution, which in turn is obtained by slightly perturbing the rest solution $\mathbf{\bar L}=\mathbf{\bar L^0}$. In other words, the elastic potential remain mono-stable along the entire relaxation. Upon release from $\mathbf{L}_\text{hold}$, the configuration will snap to $\mathbf{L}_\text{s}$ and evolve back with $\mathbf{\bar L}$ to $\mathbf{\bar L^0}$. Put in more conventional terms the relaxation of the forces during the holding was not enough to stabilize the mass near the position of relaxation and when released it snaps back to a position near its original rest position.

The second limit to consider is that of small $\a \rightarrow 0$, FIG.~\ref{fig:sls_von_mises_truss_lengths_space}c. Low values of $\a$ diminish the value of the eccentricity towards unity, shrink the hyperbola vertically and transform the two asymptotic lines to become closer to parallel. Thus, as expected, the rest lengths $\mathbf{\bar L^0}$ is now associated with two stable points; the original rest position and another stable point, $\mathbf{L}_\text{i}$, situated closer to the inverted configuration $\mathbf{L}_\text{hold}$, and associated with $z<0$. During relaxation the position of the two stable configurations also evolve slightly, $\mathbf{L}_\text{s}$ for some another $z>0$ and $\mathbf{L}_\text{i}$ for different $z<0$. The system remains bi-stable, and the relaxation only further lowers the instantaneous elastic energy of the inverted states (as their distance decreases).
	

If for this case we follow a different protocol in which the systems starts at rest and then brought abruptly to the state $\mathbf{L}=\mathbf{L_{i}^0}$ which is a locally stable equilibrium with respect to $\mathbf{\bar L}=\mathbf{\bar L^0}$, then while $\mathbf{\bar L}$ will evolve towards $\mathbf{L_{i}^0}$, the actual configuration will remain unchanged at $\mathbf{L}=\mathbf{L_{i}^0}$. We term such states stationary states. This stationarity property can be easily understood geometrically. In the protocol described  $\mathbf{\bar L}$ evolves along the line connecting $\mathbf{L_{i}^0}$ and $\mathbf{\bar{L}^0}$ for every point on this line the assumed configuration $\mathbf{L_{i}^0}$ is locally the closest to it, and thus the position of the elastic equilibrium does not change.

Last we consider the case of intermediate values of $\a$ that display delayed instability. One could find values of $\a$ large enough such that for $\mathbf{\bar L}=\mathbf{\bar L^0}$ only the state $\mathbf{L}=\mathbf{\bar L}$ is stable, yet small enough such that after relaxation at $\mathbf{L}_\text{hold}$ for time $t$ there will be two distinct locally stable equilibria with respect to $\mathbf{\bar L}(t)$, as depicted in FIG.~\ref{fig:sls_von_mises_truss_lengths_space}b. When released from relaxation the system will assume the state  $\mathbf{L}=\mathbf{L}_\text{i}$, where $z<0$. However, the reference length $\mathbf{\bar L}$ continuously evolves towards collinearity with $\mathbf{L}$ and $\mathbf{\bar L^{0}}$. However, as no state other than the trivial one is stable with respect to $\mathbf{\bar L}=\mathbf{\bar L^0}$ such a process cannot converge to a stable state and thus must at some time loose stability. 
 
The Von-Mises truss SLS models displays non-trivial phenomenology of instabilities of viscoelastic systems, all of which are intuitively explained through the graphical representation. If the system at rest can be brought to another locally stable equilibrium, then the equilibrium point will remain unchanged despite the visco-elastic evolution. In particular it will never become unstable. Conversely, state that lose stability at a finite time must have not been stable with respect to their rest state and their stability must have been acquired through relaxation. Last, every acquired stable point cannot remain stable indefinitely and must become unstable. The results obtained here were obtained for the specific geometry and constitutive relations of the Von-Mises SLS system. However, they remain valid for rather general linear visco-elastic systems. In appendix \ref{app:1d_sls_systems} we show that they hold for an arbitrary collection of SLS's provided the mass they are connected to is constrained to move along a line and they all share the same $\b$ and $\tau$. We next come to generalize these result for the general isotropic and incompressible three dimensional linearly viscoelastic continua.

\section{Metric linear viscoelasticity}
	\label{sec:metric_linear_viscoelasticity}
We now turn to discuss linear viscoelasticity of continuous bodies. Such systems are amenable to a geometric reformulation 
similar to what we did for the SLS model. Here we need to exploit an infinitesimal analog of the evolving reference lengths, which we achive by resorting to an elastic description using metric tensors. 
	\subsection{Linear viscoelastic continuum}
		\label{sec:viscoelastic_continuum}
We start by equipping the body with a material (or Lagrangian) coordinate system $x^1,x^2,x^3$. $\textbf{r}(x^i,t)$ is the configuration of the body at time $t$ in $\mathbb{R}^3$. The configuration manifest itself on the coordinate system (up to time dependent rigid motions) by the induced Euclidean metric $g_{ij}=\partial_i \textbf{r}\cdot \partial_j \textbf{r}$, which measure the infinitesimal lengths in the solid~\cite{ESK09}. In analogy with $\bar L^0$ we define the \emph{rest reference metric} $\gbaro_{ij}$ as the metric on which the body is locally stress-free and stationary. That is, if the body was to be cut to infinitesimal pieces and each of the pieces allowed to freely relax indefinitely, then the lengths of each piece will approach those represented by the rest reference metric $\gbaro_{ij}$. The strain tensor of the body is 
		\begin{equation}
			\e_{ij}(t) = \tfrac 1 2 \left(g_{ij}(t)-\gbaro_{ij} \right). \label{eq:strain_relation}
		\end{equation}
Different measures of stress correspond to different measures of volume. The second Piola-Kirchoff stress tensor $S^{ij}$ is the virtual work conjugate of the metric, per unit volume $\sqrt{\gbaro}$ ~\cite{LL86,ESK09}
		\begin{align}
		 	\d W & = -\int S^{ij} \d \e_{ij} \sqrt{\gbaro} d^3x \nn
		 	&  = -\frac{1}{2} \int S^{ij} \d g_{ij} \sqrt{\gbaro} d^3x \label{eq:stress_definition}
		\end{align}
		Intuitively, the energy difference obtained by varying only the distance between two points is exactly the force between them.
		The general viscoelastic model assumes the stress at time $t$ is a functional of the entire history of the strain $S^{ij}(t) = S^{ij}[\e_{ij}(s)]$ of times $s \le t$. $\gbaro_{ij}$ was described above as the only stationary stress-free configuration. By Eq.~\eqref{eq:strain_relation}, $S^{ij}(t)=0$ for all $t$ if and only if the strain $\e_{ij}(t)=0$ for all $t$. Following~\cite{CN61} we can approximate the functional $S^{ij}[\e_{ij}]$ around the zero function, and obtain a linear functional
		\begin{equation}
			S^{ij}(t) = C^{ijkl}\e_{kl} + \int_{-\infty}^t \Phi^{ijkl}(t-s) \e_{kl}(s)\ ds.
			\label{eq:tensor_linear_viscoelasticity}
		\end{equation}
		$C^{ijkl}$ is the stiffness tensor, $\Phi^{ijkl}(s)$ is the \emph{memory kernel} accounting for the viscoelastic dissipation. $\Phi(s)$ should fade over time, thus we take $\dot \Phi(s)<0$ and $\Phi(s\rightarrow\infty)=0$~\cite{CN61}. 

		Similar to the SLS, Eq.~\eqref{eq:tensor_linear_viscoelasticity} predicts that instantaneous incremental deformations $\Delta g_{ij}$ lead to linear stress increments $\Delta S^{ij} = \tfrac{1}{2}C^{ijkl}\Delta g_{kl}$. Following the above notion of temporal reference length $\bar L$, we define the time-dependent \emph{reference metric} $\gbar_{ij}$ of the body to satisfy
		\begin{equation}
			S^{ij}(t) = C^{ijkl}\tfrac{1}{2} \left(g_{kl}(t)- \gbar_{kl}(t) \right). \label{eq:stress_relation}
		\end{equation}
		The temporal evolution of the reference metric can be deduced from \cref{eq:scalar_tensor_linear_viscoelasticity,eq:strain_relation,,eq:stress_relation}, and reads
		\begin{align}
			\gbar_{ij}(t) & = \gbaro_{ij} - C^{-1}_{ijkl} \int_{-\infty}^{t} \dot \Phi^{klmn}(t-s)\left(g_{mn}(s) -\gbaro_{mn}\right)ds \\
			& = \gbaro_{ij} - \int_{-\infty}^{t} \dot {\hat \Phi}^{kl}_{ij}(t-s)\left(g_{kl}(s) -\gbaro_{kl}\right)\ ds. \label{eq:gbar_evolution_rule}
		\end{align}
		$\hat \Phi^{kl}_{ij}(s) = C^{-1}_{ijkl} \Phi^{klmn}(s)$ is the \emph{metric memory kernel}. As we can see, $\bar g_{ij}$ is defined only on cases where the stiffness tensor is invertible, a reasonable assumptions for most solids.
		As expected, $\gbar_{ij}(t)$ remains unchanged by instantaneous variations of $g_{ij}$. We may thus consider it as the slow state variable describing the viscoelastic evolution of the material. At each moment in time we may consider the system as an elastic system with respect to the reference metric $\gbar_{ij}$~\cite{ESK09,ESK13}.

	\subsection{The quasi static approximation}
		\label{sec:quasi_static_approximation}
		Viscoelastic systems are dissipative, thus the notion of an elastic free energy is ill defined. Nonetheless, the virtual work of spatial variations performed by the body over a short period $\Delta t \rightarrow 0 $, coincides with the instantaneous elastic energy functional~\cite{ESK09}		
		\begin{equation}
			E\left[\e^\text{el}_{ij}\right]=\int \frac{1}{2}C^{ijkl} \e^\text{el}_{ij} \e^\text{el}_{kl}\ \sqrt{|\gbaro|}\ d^3x,
			\label{eq:energy_functional}
		\end{equation}
		where the elastic strain is again $\e^\text{el}_{ij} = \tfrac{1}{2}\left(g_{ij}-\gbar_{ij}\right)$. Typically the elastic response time scale in elastomers is much smaller than the viscoelastic relaxation time. In such cases we can eliminate inertia from the system and approximate the motion of material as quasi static evolution between elastic equilibrium states. That is, the configuration at every moment in time, given by the metric $g_{ij}(t)$, minimizes the instantaneous elastic energy functional \eqref{eq:energy_functional}. If we can characterize 
		 the possible Euclidean metrics by finite set of variables $\l_\a$, i.e. $g_{ij}(x) = g_{ij} [ \l_\a ] (x)$ then the minimization condition is for all $\a$, reads $\frac{\delta E}{\delta \l_\a}=0$ or
		\begin{equation}
			F_\a\left[g_{ij}\right]=-\frac{1}{2}\int \frac{\delta g_{ij}}{\delta \l_\a} S^{ij} \ \sqrt{|\gbaro|}\ d^3x =0, \label{eq:minimaztion_condition}
		\end{equation}
		which is the vanishing of the generalized forces. For this extremal point to be a minima, the Hessian, $\frac{\d^2E}{\d \l_\a \d \l_\b}$, needs to be positive definite.

		Condition \eqref{eq:minimaztion_condition} gives $g_{ij}(t)$ as function of $\gbar_{ij}(t)$, which in turn evolve according to Eq.~\eqref{eq:gbar_evolution_rule}.
		The quasi static evolution gives an intuitive interpretation of the slow evolution of viscoelastic bodies. Over time, the material's configuration quickly minimizes its 'distance' to the reference metric. The idea that in viscoelasticity bodies adapts to their configuration is implemented by the slow dynamics of the reference metric from the rest reference metric toward the recent configurations.
		This coupled evolution of $g_{ij}$ and $\gbar_{ij}$ is also simpler to analyze both numerically and theoretically compared with the full dynamics.

	\subsection{Acquired stability and stability loss}
		\label{sec:delayed_stability_and_loss}
		In section \ref{sec:sls_von_Mises_truss} we describe how the evolution of the reference lengths in discrete, 1D systems can result in non trivial evolution of the configuration (characterized by $z$). When the parameter $\a$ is appropriately set the von-Mises truss could acquire new type of stable states; namely states that are unstable initially, but acquire stability through stress relaxation at a strained state.   
Such states also exist for continuous viscoelastic  systems, and similarly to the discrete case described above can be shown in some cases to not be able to remain stable indefinitely. 

		A given reference metric $\gbar_{ij}$ can yield multiple elastically stable configurations of the instantaneous elastic energy functional \eqref{eq:energy_functional}. Under external loading, as the reference metric slowly evolves viscoelastically according to Eq.~\eqref{eq:gbar_evolution_rule} it could acquire new stable configurations, merge existing stable points or cause stable elastic configurations to lose their stability. 
		Acquiring stability is most easily understood by examining loading at constant displacements. Consider a body taken from rest to an initially unstable configuration $g_{ij}^\text{hold}$. The body is held fixed at this position and the reference metric $\gbar_{ij}$ starts from $\gbaro_{ij}$ and slowly advances toward the metric  $g_{ij}^\text{hold}$, and the forces pushing the body away from its present state gradually relax. When releasing the body an initially unstable metrics near $g_{ij}^\text{hold}$ might display transient stability. This phenomenon is called \emph{acquired stability}.

State with acquired stability lose their stability through \emph{delayed stability loss}. Seemingly stable state gradually creep until at some point elastic stability is lost, and the body rapidly snaps to a near stable configuration. 
Both acquired stability and delayed stability loss are expected for linear viscoelastic bodies, close to multi-stability. Described quantitatively in section \ref{sec:scalar_viscoelasticity} below, as the memory kernel become comparable to the stiffness, more configurations are able to acquire stability.

	\subsection{Isotropicity and homogeneity}
	Considering isotropic materials the  stiffness and memory kernel tensors $C^{ijkl}$ and $\Phi^{ijkl}(s)$ further simplify. Isotropicity is the infinitesimal invariance to rotation in some configuration of the body. More generally, it is the infinitesimal invariance to orthogonal transformations induced by some metric on the body. This metric describes how each piece of the body should be deformed in order to have the same value under rotations.
		
		By Eq.~\eqref{eq:energy_functional}, the metric associated with the stiffness is the metric used to raise the indices of the elastic strain to create the scalar energy density.
		In linear elasticity, in which $\gbar_{ij} = \gbaro_{ij}$, the stiffness is usually taken as isotropic with respect to the reference metric $\gbar_{ij}$~\cite{ESK09}. This amounts to assuming the body is isotropic in the undeformed configuration. The natural generalization of this assumption to viscoelasticity is to take the stiffness as isotropic with respect to the instantaneous reference metric $\gbar_{ij}$. This will yield isotropic response properties about the instantaneous stress free state. However, an isotropic elastic response is often intimately related to isotropic material composition and internal structure which in our case is set by the rest reference metric $\gbaro_{ij}$. Therefore, we assume here the stiffness tensor to be isotropic with respect to the rest reference metric
		\begin{equation}
			C^{ijkl} = \l \  \gbar_0^{ij} \gbar_0^{kl} + 2\mu \  \gbar_0^{ik} \gbar_0^{jl}.
		\end{equation}
We note that in principle, some materials may display a different constitutive behaviors, and in particular be isotropic with respect the instantaneous reference metric $\gbar_{ij}$ in which case the elasticity tensor should be adapted accordingly. Moreover, while it is reasonable to assume that the memory kernel $\Phi^{ijkl}(s)$ accounting for the viscoelastic response will also be isotropic with respect to $\gbaro_{ij}$, 
there is no restriction preventing it from being isotropic with respect to any linear combination of $\gbaro_{ij}$ and $\gbaro_{ij}$.

		The assumption of homogeneity is usually obtained by taking the tensor at hand as constant in space. However, because both $C^{ijkl}$ and $\Phi^{ijkl}(s)$ are densities of properties, homogeneity is a matter of measure. 
		Directly from Eq.~\eqref{eq:energy_functional}, constant stiffness gives equal energy density per 'initial' volume element $\sqrt{\gbaro}d^3x$ (with the same deformation). Here too, in principle, one can account for other measures induced by other metrics, and thus for different types of homogeneity. Thus even when a material is considered isotropic and homogeneous, there are some choices as to how to express these symmetries in the constitutive relations.

		We now focus on the reasonable and simplest option, where the stiffness and the memory kernel are both isotropic and homogeneous with respect to $\gbaro_{ij}$. The general form of the metric memory kernel ${\hat \Phi}^{kl}_{ij}(s)$ is
		\begin{equation}
			\hat \Phi^{kl}_{ij}(s) = \b \phi(s) \d_i^k \d_j^l + \psi(s) \ \gbaro_{ij} \gbar^{kl}_0\label{eq:isotropic_memory_kernel}
		\end{equation}
		where $0<\b<1$ is defined by taking $\phi(0)=1$, and both $\phi(s),\psi(s)$ decrease monotonically to zero as $s\rightarrow\infty$. Substitution in Eq.~\eqref{eq:gbar_evolution_rule} gives the isotropic evolution of the reference metric
		\begin{equation}
			\gbar(t) = (1-\b -\a(x,t))\ \gbaro_{ij} - \b \int_{-\infty}^{t} \dot \phi(t-s)\ g_{ij}(s)\ ds, \label{eq:gbar_isotropic_evolution_rule}
		\end{equation}
		where 
		\begin{equation}
			\a(x,t) = \int_{-\infty}^t \dot \psi(t-s) \left(\gbar^{kl}_0 g_{kl}(x,s)-3\right) ds.
		\end{equation}
		Note that $\a(x,t)$ is the only term in Eq.~\eqref{eq:gbar_isotropic_evolution_rule} that mixes the indices of the metrics. 

\section{Viscoelasticity of incompressible materials}
	\label{sec:scalar_viscoelasticity}
	\subsection{Constitutive relations}
		Incompressible isotropic materials are described by Poisson ratio $\nu=\tfrac{1}{2}$. The incompressible Poissonian contraction is expected to hold also during the viscoelastic relaxation. Thus, we expect $\psi(s)=0$. 
		Put in Eq.~\eqref{eq:isotropic_memory_kernel} and Eq~\eqref{eq:tensor_linear_viscoelasticity}, we obtain the following constitutive relation between stress and strain 
		\begin{equation}
			S^{ij}(t) = C^{ijkl} \left( \e_{kl}(t) + \b \int_{-\infty}^{t} \dot \phi(t-s) \e_{kl}(s)ds \right) . \label{eq:scalar_tensor_linear_viscoelasticity}
		\end{equation}
		The incompressible evolution of the reference metric (by \cref{eq:gbar_evolution_rule,eq:isotropic_memory_kernel}) also gives
		\begin{equation}
			\gbar_{ij}(t) = (1-\b)\ \gbaro_{ij} - \b \int_{-\infty}^{t} \dot \phi(t-s)\ g_{ij}(s)\ ds. \label{eq:gbar_scalar_evolution_rule}
		\end{equation}		
The dimensionless factor $\beta$ quantifies the degree of viscoelasticity in the system. It corresponds to the asymptotic value of the fraction of stress relaxed
in a constant displacement experiment starting from rest. Imposing a stationary solution for $\gbar_{ij}$ and $g_{ij}$ in Eq.~\eqref{eq:gbar_scalar_evolution_rule}, we obtain
\begin{equation}
			\gbar^\text{stat}_{ij} = \b g_{ij} + (1-\b)\gbaro_{ij}. \label{eq:gbar_stationary_condition}
\end{equation}
Here $\beta$ is shown to control the degree to which the reference metric of a viscoelastic body will approach the current configuration metric; 
$\b=0$ describes a purely elastic material with a constant reference metric, and $\b=1$ describes a viscoelastic fluid in which no information is conserved indefinitely, asymptotically all stresses are relaxed, and the notion of the rest reference metric $\gbaro_{ij}$ has no meaning.

		As seen from Eq.~\eqref{eq:gbar_stationary_condition}, the long time stability of a body depends only on $\b$, and not on the specific memory kernel. It is thus instructive to discuss the SLS model.
		The memory kernel for the SLS model is given by $\phi(s)=e^{-\nicefrac{s}{\tau}}$. In analogy to Eq.~\eqref{eq:sls_L_bar_DE}, the differential equation for the reference metric reads
		\begin{align}
			\dot \gbar_{ij}(t) =& -\frac{1}{\tau}\left(\b \left(\gbar_{ij}(t) - g_{ij}(t) \right) +(1-\b) \left(\gbar_{ij}(t) - \gbaro_{ij} \right)\right) \nn
			=&-\frac{1}{\tau}\left( \gbar_{ij}(t)-\gbar^\text{stat}_{ij} (t)\right).
			\label{eq:gbar_SLS_DE}
		\end{align}
	
	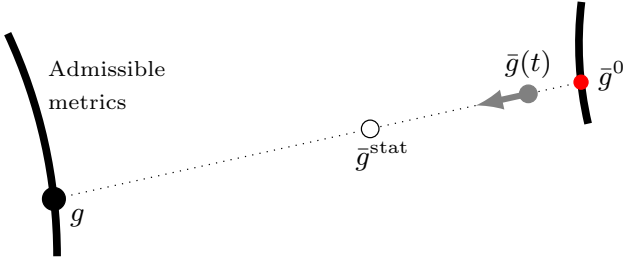
\begin{figure}[t]
		\resizebox{\linewidth}{!}{\import{.}{metric_space.tex}}
		\caption{A schematic representation of the metrics collinearity. The  minimization of the metric $g$ is constrained and performed with respect to a subset of metrics that correspond to realizable configurations. Such metrics are in particular orientation preserving and Euclidean. This set is represented by the dashed line above. Given a reference metric $\gbar$ the realized metric will correspond to the closest point from the set of admissible metrics to $\gbar$ according to the distance function given by the instantaneous elastic energy \eqref{eq:energy_functional}. Starting from rest, $\gbar$ evolves from $\gbaro$ towards $g$. Conversely, if $g$ is the closest admissible metric to $\gbar^{\text stat}$ it is also such for $\gbaro$ which is collinear with the two. 
			}\label{fig:metric_space}
	\end{figure}

	\subsection{Stationary states and the rest system}
		\label{sec:stationary_states}
		We now turn to examine the stationary solutions of the incompressible evolution rules Eq.~\eqref{eq:gbar_scalar_evolution_rule} and\eqref{eq:minimaztion_condition}. As described on section \ref{sec:sls_von_Mises_truss} and in appendix \ref{app:1d_sls_systems}, the stationary states of one dimensional SLS structures are extremal solutions of the rest elastic system $\bar L_i = \bar L_i^0$. Similar claim can be made for general linear viscoelastic and incompressible bodies. By Eq~\eqref{eq:gbar_stationary_condition} at stationarity $\gbar^\text{stat}_{ij} = \b g_{ij} + (1-\b)\gbaro_{ij}$. Configuration $g_{ij}$ is stationary if it is in elastic equilibrium with respect to its stationary reference metric $\gbar_{ij}^\text{stat}$. Thus, the generalized forces with respect to $\gbar_{ij}^\text{stat}$ must vanish
		\begin{align*}
			0 & = F^\text{stat}_\a\left[g_{ij}\right] \\
			& = -\int C^{ijkl} \frac{\d g_{ij}}{\d \l_\a} (g_{kl}-\gbar^\text{stat}_{kl}) \sqrt{|\gbaro|}\ d^3x \\
			& = -(1-\b) \ \int C^{ijkl} \frac{\d g_{ij}}{\d \l_\a} (g_{kl}-\gbaro_{kl}) \sqrt{|\gbaro|}\ d^3x\\
			& = (1-\b) F_\a^0\left[g_{ij}\right].
		\end{align*}
		$F_\a^0\left[g_{ij}\right]$ is the generalized force calculated in the rest elastic system $\gbar_{ij} = \gbaro_{ij}$. Therefore the configuration is an extremal of the elastic energy functional with respect to the rest reference metric $F^0_\a\left[g_{ij}\right]=0$. We thus obtained that all the stationary configurations of viscoelastic body are extremals of its the rest elastic system.

		Following Eqs.~\eqref{eq:energy_functional} and \eqref{eq:gbar_stationary_condition}, we arrive to the following relation between the energy Hessians of the stationary and rest systems
		\begin{equation}
			\frac{\d^{2}E_\text{stat}}{\d\l_{\a}\d\l_{\b}} = (1-\b) \frac{\d^{2}E_0}{\d\l_{\a}\d\l_{\b}} 
			+ \frac{\b}{4} \int C^{ijkl}\frac{\d g_{ij}\left(x\right)}{\d\l_{\a}}\frac{\d g_{kl}\left(x\right)}{\d\l_{\b}}
			\label{eq:Hessians_static_rest}
		\end{equation}
		Drucker stability criterion states that $C^{ijkl}$ is positive definite, thus also $\int C^{ijkl}\frac{\d g_{ij}\left(x\right)}{\d\l_{\a}}\frac{\d g_{kl}\left(x\right)}{\d\l_{\b}}$ is a positive definite matrix \cite{AB09}. If the configuration was initially stable and $\frac{\d^{2}E_0}{\d\l_{\a}\d\l_{\b}}$ is positive definite, then the stationary Hessian is also positive definite (sum of two positive definite matrices). Thus initially stable (not only extremal) configuration is also stationary. 
		But Eq.~\eqref{eq:Hessians_static_rest} also open the possibility for non-stable but extremal configuration of the rest system, for example a saddle point, to stabilize indefinitely.
		Eq.~\eqref{eq:Hessians_static_rest} also elucidate the basic intuition that when a system is held at constant displacement, stress relaxation stabilizes the configuration; as we can see from the equation, originally non positive Hessian become 'more positive' along the relaxation.
		Both features depend on the value of $\b$. As $\b\rightarrow 0$ the system almost doesn't change at all under relaxation and only initially stable configurations are also stationary. Conversely, as $\b\rightarrow 1$ the system is able to stabilize more and more configurations, and all the rest system extremals will become also  stationary configurations.
		\newline

		The second phenomenon described in section \ref{sec:sls_von_Mises_truss} and in appendix \ref{app:1d_sls_systems} deals with bringing a system to a metastable state abruptly from rest. Here we take body at rest and deform it fast into some other local elastically stable configuration. We claim that the dynamics of the system's configuration is trivial, i.e. a configuration itself will display no temporal variation despite the continuous evolution of the reference metric and the relaxation of the corresponding stress~\footnote{For the case of elasto-plasticity, a similar idea was first proved but never published by H. Aharony. This type of behavior was also conjecture~\cite{BH78}.}. First, we assume that the system starts at rest, i.e. for all $t<0$ we have $g_{ij}(t)=\gbar_{ij}(t)=\gbaro_{ij}$.  At $t=0$ the system is brought abruptly to an elastically stable conifguration
		$g_{ij}(0)\ne \gbaro_{ij}$, such that $g_{ij}(0)$ satisfies \eqref{eq:minimaztion_condition}. We wish to show that $\dot g_{ij}(t) =0$ satisfies the evolution rules for all $t>0$. First, substitution in Eq.~\eqref{eq:gbar_scalar_evolution_rule} gives
		\begin{equation}
			\gbar_{ij}(t) = \gbaro+\b\left(1-\phi\left(t\right)\right)\left(g_{ij}-\gbaro_{ij}\right), \label{eq:gbar_sol}
		\end{equation}
		where $g_{ij}=g_{ij}(t)$ is constant in time.
		More importantly, we have the evolution of the elastic strain $\e^\text{el}_{ij}(t) = \tfrac{1}{2}\left(g_{ij}-\gbar_{ij}(t)\right)$
		\begin{align}
			\e^\text{el}_{ij}(t) & = \left(1-\b\left(1-\phi\left(t\right)\right)\right) \tfrac{1}{2}\left(g_{ij}-\gbaro_{ij}\right) \nn
			& = \left(1-\b + \b \phi(t)\right) \e^\text{el}_{ij}(t=0) \label{eq:helpful}
		\end{align}
		It is left to prove that the constant $g_{ij}$ remains at local minimum of the instantaneous energy functional with respect to Eq.~\eqref{eq:gbar_sol}. First, the generalized force
		\begin{align}
			F_\a(t) & = -\frac{1}{2}\int C^{ijkl} \frac{\d g_{ij}}{\d \l_\a} \e^\text{el}_{kl}(t) \sqrt{|\gbaro|}\ d^3x \nn
			& = \left(1-\b + \b \phi(t)\right) \ F_\a(t=0) \nn
			& = 0 \nonumber
		\end{align}
		The last step uses the stability of $g_{ij}$ at $t=0$ (with respect to $\gbaro_{ij}$). Second, the Hessian 
		\begin{align*}
			\frac{\d^{2}E(t) }{\d\l_{\a}\d\l_{\b}}& = \left(1-\b + \b \phi(t)\right) \frac{\d^{2}E(t=0)}{\d\l_{\a}\d\l_{\b}} \\
			& + \frac{\b}{4} \left(1-\phi(t)\right) \int C^{ijkl}\frac{\d g_{ij}\left(x\right)}{\d\l_{\a}}\frac{\d g_{kl}\left(x\right)}{\d\l_{\b}} 
		\end{align*}
		By assumption, $\frac{\d^{2}E(t=0)}{\d\l_{\a}\d\l_{\b}}$ is a positive definite matrix. As we note above, the right matrix is also positive definite. $\frac{\d^{2}E(t)}{\d\l_{\a}\d\l_{\b}}$ is thus a sum of two positive definite matrices, and therefore also positive definite for all $t>0$. Thus, we have shown that $\dot g_{ij}(t) =0$ for any $t>0$ is the solution to the quasi static evolution equations.

		We note that the claims can not be derived under the general memory kernel $\Phi^{ijkl}(s)$ but only under the assumption that the metric kernel is scalar $\hat \Phi_{ij}^{kl} = \b \phi(s) \d_i^k \d_j^l$. Still, the result are valid to any incompressible solid in a continuous manner. That is, the actual variation from these result diminish as the material at hand is closer to incompressibility.

		Intuitively, both claims are due to collinearity of $g_{ij}$,
		$\gbaro_{ij}$ and $\gbar_{ij}$ under the appropriate initial condition (see FIG.~\ref{fig:metric_space}). If the evolution of the reference metric starts at rest $\gbar_{ij}=\gbaro_{ij}$ then evolving toward $g^\text{stat}_{ij}$ also evolve in the direction of $g_{ij}$. If the metric was originally stable - with 'minimal length' to $\gbar_{ij}=\gbaro_{ij}$ , than it will also be minimal to any reference metric $\gbar_{ij}(t)$ on the line between the rest reference metric to $g_{ij}$.

	\subsection{Transient elastic stability and snap-through} 
		\label{sec:transient_stability_loss}
\begin{figure}[t]
			\centering	
			\begin{minipage}{\linewidth}
				\subfloat[]{
					\label{fig:hoppers}
					\includegraphics[clip,trim=0cm -4.8cm 0cm 0cm,width=.38\linewidth]{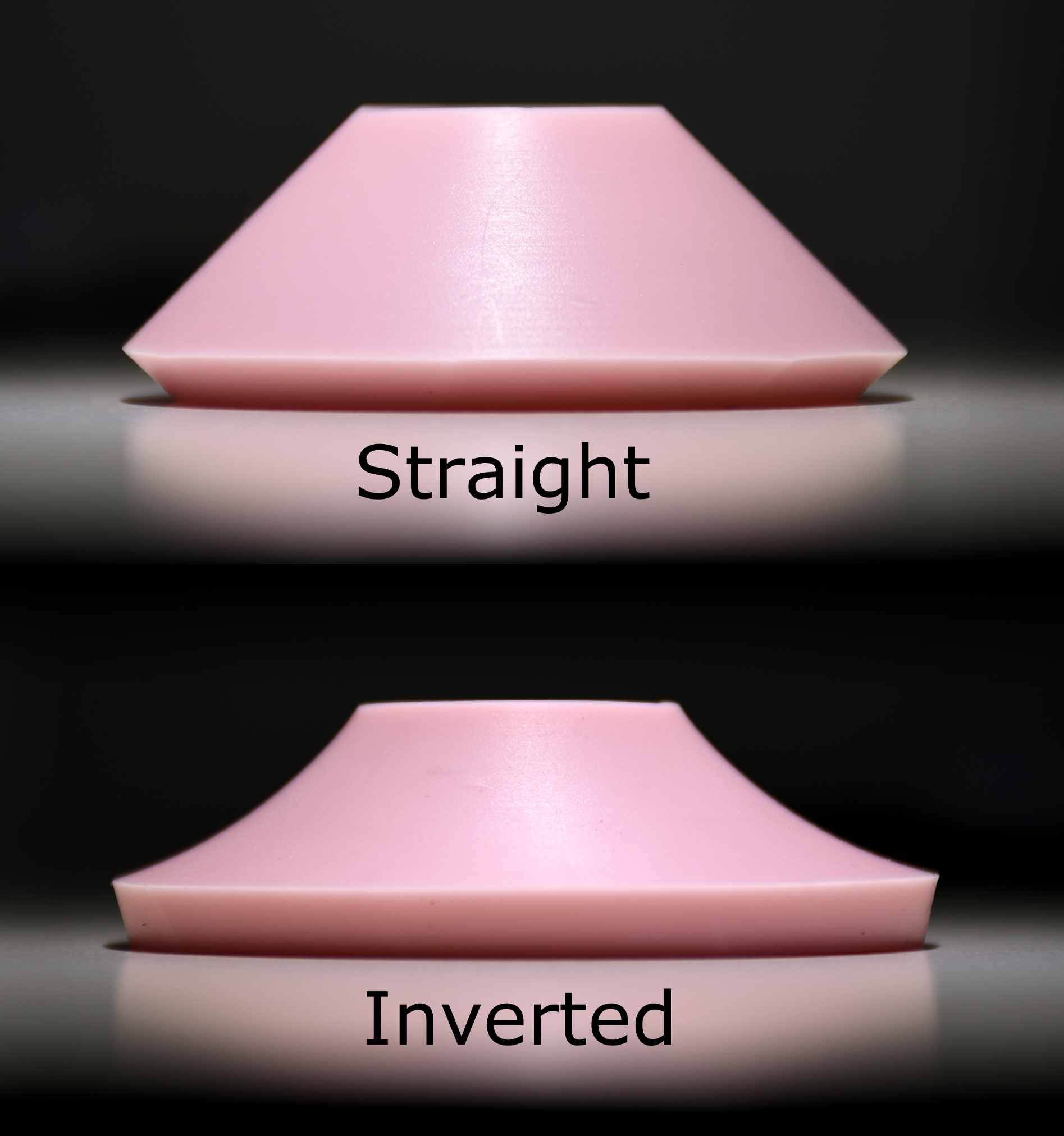}
				}
				\subfloat[]{
					\label{fig:protocol}
					\resizebox{.63\linewidth}{!}{\import{}{./protocol.pgf}}
				}
				\newline
				\subfloat[]{
					\label{fig:flip_time_vs_thickness}
					\resizebox{\linewidth}{!}{\import{}{flip_time_vs_thickness.pgf}}
				}

			\end{minipage}
			\caption{
			(a) Straight and inverted poppers.
			(b) The experiment protocol: The cone is held in the inverted shape for a time $t_\text{hold}$ and then released; it flips after a time $t_\text{flip}$.
			(c) The flipping time as function of the cone thickness for the two limiting cases of immediate release and long holding time. The elastic cone was simulated with geometrical properties $r_\text{min}=10[\text{mm}],r_{max}=25[\text{mm}]$. The material properties taken were $\b = 0.3$, Young's modulus $E=2.5[\text{MPa}]$ and Poisson ratio $\nu=0.47$. A similar phase plot presented in \cite{BSP12} and in \cite{UE18a}.
			}
			\label{fig:hopper_popper}
		\end{figure}

		We have so far discussed the behavior of viscoelastic systems in states that are stable when arrived to from rest. We next come to discuss the complementary states that are unstable. In particular we would like to address how could they acquire stability and also in turn how this stability is lost. Jumping poppers are thin rubber shells. Very thin poppers display almost isometric bi-stability, while thick poppers have only one elastically stable shape. However, when an appropriately cut popper is flipped inside out and laid on the table [Fig.~\eqref{fig:hopper_popper}] it creeps for a few seconds and snaps back to its original shape, much in the same manner systems with intermediate values of $\a$ in the Von-Mises truss model of section \ref{sec:sls_von_Mises_truss} acquired and lost stability.
				
		One general corollary of the previous section is that stability loss can occur only from configurations that were ``originally'' unstable - not a stable configuration of the rest elastic system $\gbar_{ij}=\gbaro_{ij}$. If the configuration was originally stable, than by the above claim it is also stationary, and its stability can never be lost. Thus a system exhibiting temporary stability followed by a snap though, was necessarily away of its rest system $\gbar_{ij}\ne\gbaro_{ij}$ before the free evolution. Therefore if a popper snap a finite time after being inverted, its stability (prior to the snap) was necessarily acquired, presumably by holding it inverted for a long enough duration. 
		
		Second, acquired stability from configurations that were not extremal initially (e.g. saddle points) cannot persist indefinitely and must be eventually lost. If a configuration is stationary after a relaxation that started from rest then it must also be extremal with respect to the rest reference metric. In the case of the popper it means that if it was originally unstable, there is no way (by holdings the popper inverted or by any other deformations) to stabilize the popper indefinitely after the external loads are removed. For such cases the inverted shapes cannot be stationary states of the system, and the popper will always eventually snap back to an rest state.

	\label{sec:experimental_results}

%
%
%
%
%
\section{Summary and discussion}
Naively one expects stress relaxation to only diminish internal forces and thus act to stabilize a system rendering the notion of viscoelastic instabilities elusive. We have shown this intuition to indeed be correct; if a system is brought abruptly from its rest state into an elastically meta-stable state then stress relaxation will never cause it to lose stability. We moreover proved that in such a setting the configuration of the system will show no evolution despite the continuous vicoelastic attenuation of internal stresses. 

Having formulated the viscoelastic evolution in terms of metric tensors we observed that the instantaneous reference metric evolves not only toward $g$, the assumed configuration, but also toward $\gbaro$, its rest metric. When starting from rest only the former is observed. In this case the reference metric evolves along a straight line and as a result the elastic energy minimizer at time $t$ with respect to $\gbar(t)$ remains the minimizer for subsequent times, as explained graphically in FIG.~\eqref{fig:metric_space}. For every other case, in which $\gbar$ does not evolve solely toward $g$, we expect a visible creep and in particular such systems may display viscoelastic instabilities. We have shown that the only states that can display viscoelastic instabilities are states that are elastically unstable with respect to $\gbaro$ but have acquired stability through stress relaxation. We have also shown that every instance of such an acquired stability cannot persist indefinitely and must be lost in time. 

These results are expected to hold for every isotropic and incompressible linearly viscoelastic material regardless of the form of history dependence of the stress relaxation kernel. However, considering a single SLS system or any other purely one dimensional viscoelastic system we do not observe this rich phenomenology of instabilities due to the trivial structure of the set of admissible states. The ability of a viscoelastic system to display delayed instabilities relies on the non-trivial shape of the set of admissible states. As observed in FIG.~\eqref{fig:sls_von_mises_truss_lengths_space} for the SLS Von-Misses truss, only when the set of admissible states is sufficiently curved can it support multi-stability.  For continuous media the structure of admissible (Euclidean) states is non-convex. As a result we may bring a systems that is initially at rest in an Euclidean configuration, to another Euclidean state by the application of external forces, and the viscoelastic evolution will result in a non-Euclidean instantaneous reference metric, giving rise to a geometrically frustrated state. In such cases residual stress is like to appear even in lieu of multistability. 

While in the theory presented here the rest metric, $\gbaro_{ij}$, is assumed constant one could consider growing tissue or plastically deforming material by allowing the rest metric to vary in time. The variation could be provided by external conditions or obey some additional constitutive relations. This will result in a covariant elaso-visco-plastic theory. Such a description may provide crucial insight into growing and deforming biological polymers where a rapid growth or shrinkage events may be masked by the slow evolution of the instantaneous reference metric. The applicability of such a theory, reminiscent of the additive decomposition of strains \cite{EL69,ESK13}, to growing and deforming tissue remains to shown.

%

\begin{acknowledgments}
The authors would like to thank D. Biron, G. Cohen, A. Grosberg, S.M. Rubinstein, E. Sharon, Y. Bar-Sinai and D. Vella for helpful discussions, as well as acknowledge the Aspen Center for Physics, which is supported by National Science Foundation grant PHY-1607761, for hosting many of these discussions. 
\end{acknowledgments}

\bibliography{ourBIB}

\appendix*

\section{1D SLS truss and rest state stationarity}
		\label{app:1d_sls_systems}
Here we examine an explicitly solvable discrete system that is more complex than the three spring truss. This system is composed of a point mass $m$, free to move along the $\hat z$ axis whose position we denote $z(t)$. The mass is connected through SLS's to $N$ pinned positions in the 2D space $(x_i,z_i)$. All the SLS's share the same $\b,\tau$ but may differ in their stiffnesses $k_i$. 
		This choice is well justified when considering large structures, or composite materials, made of many identical SLS's. The compositions of identical SLS's (in parallel or in series) leave $\b$ and $\tau$ unchanged. Thus the effective description of an SLS structure can assign different stiffnesses to distinct modes of deformations, but take $\b$ and $\tau$ as intrinsic material properties.
		The length on each of the SLS's is denoted $L_i = \sqrt{x_i^2 + (z_i-z)^2}$, and the reference length and rest length of each SLS is denoted $\bar L_i$ and $\bar L^0_i$ respectively.

		We assume a temporal scale separation between the fast elastic time-scale and the slow viscoelastic time shared by all the springs, and that no external time scale enters the problem. We may thus reduce the problem to studying the quasi static evolution of the system in which every configuration the system assumes minimizes the instantaneous elastic energy that in turn depends on the slowly evolving reference lengths $\bar L_i$,
		\begin{equation}
			E(z) = \sum_{i=1}^N \frac{1}{2} k_i \left(L_i-\bar L_i\right)^2.
		\end{equation}	
In this quasi-static limit the force on the point mass vanishes
	 	\begin{align}
	 		0 = F(z) & = -\frac{d E}{d z} = -\sum_{i=1}^N k_i \frac{dL_i}{dz} \left(L_i-\bar L_i\right). \label{eq:sls_1d_force}
	 	\end{align}
The rest length $\bar L_i$ change in time and, as in the case of the Von-Mises truss, could change the shape of the instantaneous elastic energy. For example they could cause an unstable state to become stable through viscoelastic relaxation. Such states with acquired stability in viscoelastic systems were first identified in the context of creep buckling \cite{NH56,BH78}, and revisited recently where this phenomenon  was termed temporary bistability \cite{MS10} or pseudo bistability \cite{BSP12}, as these states were observed numerically to always lose their stability. 

	 	In order to analyze the dynamic of the system, we differentiate the force Eq.~\eqref{eq:sls_1d_force}  in time, and use Eq.~\eqref{eq:sls_L_bar_DE}: 
	 	\begin{align}
	 		\frac{d^2 E}{dz^2} \dot z & = \sum_{i=1}^N k_i \frac{dL_i}{dz} \dot {\bar L}_i \nn
	 		& = \frac{1}{\tau} \sum_{i=1}^N k_i \frac{dL_i}{dz} 
	 		\left( L_i - \bar L_i - (1-\b)\left(L_i-\bar L_i ^0\right)\right) \nonumber
	 	\end{align}
	 	Substitution of Eq.~\eqref{eq:sls_1d_force} gives
	 	\begin{equation}
	 		\frac{d^2 E}{dz^2} \dot z = \frac{1-\b}{\tau} F^0(z),
	 		\label{eq:sls_1d_quasistatic_dynamic}
	 	\end{equation}
	 	where $F^0(z)=-\sum_{i=1}^N k_i \frac{dL_i}{dz} \left(L_i-\bar L_i^0\right)$ is the force on the mass at $z$, with respect to reference lengths that equal the rest lengths, $\bar L_i = \bar L_i^0$.
	 	We thus obtained that when the point mass is in a convex region of the potential ($\frac{d^2 E}{dz^2}>0$), $z$ changes continuously and the sign of $\dot z$ remains the same as the sign of the force in the rest elastic system $F^0(z)$. When the second derivative vanishes $\frac{d^2 E}{dz^2}=0$ the system loses its transient elastic stability, and snaps to another stable point.

	 	Immediate result of Eq.~\eqref{eq:sls_1d_quasistatic_dynamic} is that stationary states of Eqs.~\eqref{eq:sls_1d_quasistatic_dynamic} and \eqref{eq:sls_L_bar_DE} are all extremal points of the rest elastic system's potential. That is assuming both $\dot z =0$ and $\dot{\bar L}_i=0$ implies $F^0(z)=0$. 
	 	This result agrees with the phenomenology of the von-Mises truss described above. Both the unstable and the finitely stable cases eventually result in the original position - stable point of the rest elastic system. The stable phase satisfy the claim trivially, by staying stable from the very rest. We have thus also shown that in general, any acquired stability (initially non extremal point that has become stable), will eventually result in a snap \footnote{Technically, another possibility is that the point will continuously progress to one of the rest elastic system extremum points.}.

	 	The sign of $\dot z$ in Eq.~\eqref{eq:sls_1d_quasistatic_dynamic} is independent of the reference lengths. Thus, mass left on a stable point of the rest elastic system $F^0(z)=0$ will remain static, although all the reference length $\bar L_i$ will continue to change. 
	 	Intuitively, this phenomenon occurs because the relaxation takes the same force fraction of each SLS over time, preserving the vanishing total force on the point mass. One only need to check whether $\frac{d^2 E(t)}{dz^2}>0$ during the relaxation. The second derivative reads
	 	\begin{equation}
	 		\frac{d^2 E(t)}{dz^2} = \sum_{i=1}^N k_i \frac{d^2L_i}{dz^2} \left(L_i-\bar L_i(t)\right) + \sum_{i=1}^N k_i \left(\frac{dL_i}{dz}\right)^2
	 	\end{equation}
	 	Assuming the point to be at rest $\bar L_i = \bar L_i^0$ at $t=0$, and taking constant $z$ and $L_i$'s, Eq.~\eqref{eq:sls_L_bar_DE} gives 
		\[\bar L_i(t) = \left( 1- B(t)\right) \bar L_i^0 + B(t) L_i,\]
	 where $B(t)= \b \left(1-e^{-\nicefrac t \tau}\right)$. Thus
	 	\begin{equation}
	 		\frac{d^2 E(t)}{dz^2} = \left( 1- B(t)\right) \frac{d^2 E(t=0)}{dz^2} + B(t) \sum_{i=1}^N k_i \left(\frac{dL_i}{dz}\right)^2
	 	\end{equation}
	 	Because $0<B(t)<1$ for all $t>0$ we get that if the position was stable at rest $\frac{d^2 E(t=0)}{dz^2}>0$ it would indeed remain stable for any $t>0$. We have therefore obtained that stable points of the rest elastic system are also static, when arrived to directly from rest. 
		
By considering the SLS as a spring with dynamic reference length we were able to show explicitly that the results obtained for the Von-Mises truss SLS system are in fact rather general and in particular apply even when the system display multiple metastable states and is comprised of many springs of different stiffnesses provided their $\b$ and $\tau$ values coincide.  
%

\end{document}

%% file: SLS_diagram.tex
\def\len{.1}
\def\ld_dx{0.3}
\def\l_dx{0.2}
\begin{tikzpicture}[scale=1.4]
	\tikzstyle{spring}=[thick,decorate,decoration={zigzag,pre length=0.3cm,post length=0.3cm,segment length=6}]
	\tikzstyle{damper}=[thick,decoration={markings,  
	  mark connection node=dmp,
	  mark=at position 0.5 with 
	  {
	    \node (dmp) [thick,inner sep=0pt,transform shape,rotate=-90,minimum width=15pt,minimum height=2pt,draw=none] {};
	    \draw [thick] ($(dmp.north east)+(2pt,0)$) -- (dmp.south east) -- (dmp.south west) -- ($(dmp.north west)+(2pt,0)$);
	    \draw [thick] ($(dmp.north)+(0,-5pt)$) -- ($(dmp.north)+(0,5pt)$);
	  }
	}, decorate]
	\tikzstyle{ground}=[fill,pattern=north east lines,draw=none,minimum width=0.75cm,minimum height=0.3cm]
	\tikzset{dim/.style args={#1,#2}{decoration={add dim,distance=#2},
	                decorate,
	                postaction={decorate,decoration={text along path,
	                                                 raise=#2,
	                                                 text align={align=center},
	                                                 text={#1}}}}}

	\draw [spring] (.5,2.5) -- (.5,.5);
	\draw [spring] (-.5,2) -- (-.5,.5);
	\draw [damper] (-.5,2.5) -- (-.5,2);

	\draw[thick] (0,3) -- (0,2.5) (-.5,2.5)--(.5,2.5) (-.5,.5)--(.5,.5) (0,.5) -- (0,0);

	\draw ({.5+\l_dx},2.1) node[right] {\scriptsize $(1-\beta)k$};

	\draw (-.6,1.25) node[left] {\scriptsize $\beta  k$};
	\draw (-.2,2.25) node[right] {\scriptsize $\eta$};

	\draw ({.5+\l_dx-\len},0.5) -- ({.5+\l_dx+\len},.5) -- ({.5+\l_dx},0.5) -- ({.5+\l_dx},2.5) -- ({.5+\l_dx-\len},2.5) -- ({.5+\l_dx+\len},2.5);
	\draw ({.5+\l_dx},1.5) node[right] {\scriptsize $L$};

	\draw ({-.5-\ld_dx-\len},2.5) -- ({-.5-\ld_dx+\len},2.5) -- (-.5-\ld_dx,2.5) -- (-.5-\ld_dx,2) -- ({-.5-\ld_dx-\len},2) -- ({-.5-\ld_dx+\len},2);
	\draw (-.5-\ld_dx,2.25) node[left] {$\scriptsize L_d$};

\end{tikzpicture}

%% file: SLS_model_diagram.tex
\def\len{2}
\def\extra{0.3}
\def\sls_width{7}
\def\scale{1}
\begin{tikzpicture}[scale=\scale]
	\tikzstyle{spring}=[thick,decorate,decoration={zigzag,pre length=0.1cm,post length=0.1cm,segment length=6}]
	\tikzstyle{damper}=[thick,decoration={markings,  
	  mark connection node=dmp,
	  mark=at position 0.5 with 
	  {
	    \node (dmp) [thick,inner sep=0pt,transform shape,rotate=-90,minimum width=15pt,minimum height=2pt,draw=none] {};
	    \draw [thick] ($(dmp.north east)+(2pt,0)$) -- (dmp.south east) -- (dmp.south west) -- ($(dmp.north west)+(2pt,0)$);
	    \draw [thick] ($(dmp.north)+(0,-5pt)$) -- ($(dmp.north)+(0,5pt)$);
	  }
	}, decorate]
	\tikzstyle{ground}=[fill,pattern=north east lines,draw=none,minimum width=0.75cm,minimum height=0.3cm]
	\tikzstyle{sls}=[opacity=0,thick,decoration={markings,  
	  mark=at position 0 with 
	  {
	  	\node (sls_start) {};
	  }
	  mark connection node=sls_west,
	  mark=at position 0.2 with 
	  {
	  	\node (sls_west) {};
	  },
	  mark=at position 0.5 with 
	  {
	  	\node (sls_mid) {};
	  },
	  mark=at position 0.8 with 
	  {
	  	\node (sls_east) {};
	  	\draw [spring] ($(sls_west)+(0,{\sls_width pt})$) -- ($(sls_east)+(0,{\sls_width pt})$);
	  	\draw [spring] ($(sls_west)+(0,{-\sls_width pt})$) -- ($(sls_mid)+(0,{-\sls_width pt})$);
	  	\draw [damper] ($(sls_mid)+(0,{-\sls_width pt})$) -- ($(sls_east)+(0,{-\sls_width pt})$);
	  	\draw [black,thick] ($(sls_west)+(0,{\sls_width pt})$) -- ($(sls_west)+(0,{-\sls_width pt})$);
	  	\draw [black,thick] ($(sls_east)+(0,{\sls_width pt})$) -- ($(sls_east)+(0,{-\sls_width pt})$);
	  },
	  mark=at position 1 with 
	  {
	  	\node (sls_end) [minimum width=0pt,minimum height=0pt] {};
	  	\draw[black,thick] ($(sls_start)$)--($(sls_west)+(0,0.pt)$) ($(sls_east)+(0,-0.pt)$) --($(sls_end)$);
	  }
	}, decorate]

	\draw (0,\len) node[ground, minimum width=2cm,anchor=south](sls_up) {};
	\draw (sls_up.south east) -- (sls_up.south west);

	\draw ({(\len/sqrt(2)+\extra)},{-(\len/sqrt(2)+\extra)}) node(sls_right){};
	\draw (sls_right) node[ground, minimum width=.5cm,anchor=south,rotate around={90:(sls_right)}](sls_right_wall) {};
	\draw (sls_right_wall.north east) -- (sls_right_wall.north west);

	\draw ({-(\len/sqrt(2)+\extra)},{-(\len/sqrt(2)+\extra)}) node(sls_left){};
	\draw (sls_left) node[ground, minimum width=.5cm,anchor=south,rotate around={-90:(sls_left)}](sls_left_wall) {};
	\draw (sls_left_wall.north east) -- (sls_left_wall.north west);

	\draw [sls] (0,0) -- (sls_up);
	\draw [sls] (0,0) -- (sls_right_wall.north);
	\draw [sls] (0,0) -- (sls_left_wall.north);

	\draw[black,right] ({\len/sqrt(2)*0.4},{-\len/sqrt(2)*0.4}) node[right=15pt] {\scriptsize $k_1$};
	\draw[black] (0,{\len/2}) node[right={\sls_width + 7}] {\scriptsize $k_2$};

\end{tikzpicture}

%% file: metric_space.tex
\newcommand*{\pgfmathsetnewmacro}[2]{%
    \newcommand*{#1}{}
    \pgfmathsetmacro{#1}{#2}%
}

\begin{tikzpicture}[scale=2]
	\pgfmathsetnewmacro{\len}{3}
	\pgfmathsetnewmacro{\rad}{3}
	\pgfmathsetnewmacro{\ang}{25}
	\pgfmathsetnewmacro{\gang}{\ang*.25}
	\pgfmathsetnewmacro{\circlerad}{.05}
	\pgfmathsetnewmacro{\m}{tan(\ang/2)}
	\pgfmathsetnewmacro{\xo}{\rad*cos(\gang)}
	\pgfmathsetnewmacro{\yo}{{\rad*sin(\gang)}}

	\pgfmathsetnewmacro{\rado}{2}
	\pgfmathsetnewmacro{\ango}{20}
	\pgfmathsetnewmacro{\xoo}{\xo+\len+\rado*cos(\gang)+\rado*cos(180+(\gang-\ango)*.5)}
	\pgfmathsetnewmacro{\yoo}{\yo+\m*\len+\rado*sin(\gang)+\rado*sin(180+(\gang-\ango)*.5)}

	\draw (\xo+.6*\len,\yo+\m*.6*\len) node(gstat){} circle[radius=\circlerad];

	\draw[line width=2.5] (\rad,0) arc(0:\ang:\rad)
	(\xoo, \yoo) arc(180+(\gang-\ango)*.5:180+(\gang+\ango)*.5:\rado);

	\draw[fill=black] (\xo,\yo) node(g){} circle[radius=\circlerad*1.3];
	\draw[fill=red,color=red] (\xo+\len,\yo+\m*\len) node(gbar0){} 
			circle[radius=\circlerad*.8];
	\draw[color=black!50,fill=black!50] (\xo+.9*\len,\yo+\m*.9*\len) node(gbar){} 
							(gbar.center) circle[radius=\circlerad];

	\draw[dotted] (g) node[below=6pt,right=2pt]{$g$} 
	-- (gstat) node[right=4pt,below=1pt]
	{$\bar g^{\text{stat}}$}
	-- (gbar) -- (gbar0) node[right=10pt,below=-10pt]{$\bar g^0$};
	\draw (gbar.center) node[above=2pt]{$\bar g(t)$};
	\draw[-latex,line width=2,color=black!50] (gbar.center) -- (\xo+.8*\len,\yo+\m*.8*\len);
	
	\draw[anchor =south west] ({\xo*.97},{\yo*2.4}) node[text width=1.5cm]
	{\scriptsize Admissible metrics};
\end{tikzpicture}

%% file: protocol.pgf
\begingroup%
\makeatletter%
\begin{pgfpicture}%
\pgfpathrectangle{\pgfpointorigin}{\pgfqpoint{4.000000in}{3.500000in}}%
\pgfusepath{use as bounding box, clip}%
\begin{pgfscope}%
\pgfsetbuttcap%
\pgfsetmiterjoin%
\definecolor{currentfill}{rgb}{1.000000,1.000000,1.000000}%
\pgfsetfillcolor{currentfill}%
\pgfsetlinewidth{0.000000pt}%
\definecolor{currentstroke}{rgb}{1.000000,1.000000,1.000000}%
\pgfsetstrokecolor{currentstroke}%
\pgfsetdash{}{0pt}%
\pgfpathmoveto{\pgfqpoint{0.000000in}{0.000000in}}%
\pgfpathlineto{\pgfqpoint{4.000000in}{0.000000in}}%
\pgfpathlineto{\pgfqpoint{4.000000in}{3.500000in}}%
\pgfpathlineto{\pgfqpoint{0.000000in}{3.500000in}}%
\pgfpathclose%
\pgfusepath{fill}%
\end{pgfscope}%
\begin{pgfscope}%
\pgfsetbuttcap%
\pgfsetmiterjoin%
\definecolor{currentfill}{rgb}{1.000000,1.000000,1.000000}%
\pgfsetfillcolor{currentfill}%
\pgfsetlinewidth{0.000000pt}%
\definecolor{currentstroke}{rgb}{0.000000,0.000000,0.000000}%
\pgfsetstrokecolor{currentstroke}%
\pgfsetstrokeopacity{0.000000}%
\pgfsetdash{}{0pt}%
\pgfpathmoveto{\pgfqpoint{0.407116in}{0.745593in}}%
\pgfpathlineto{\pgfqpoint{3.647572in}{0.745593in}}%
\pgfpathlineto{\pgfqpoint{3.647572in}{3.263907in}}%
\pgfpathlineto{\pgfqpoint{0.407116in}{3.263907in}}%
\pgfpathclose%
\pgfusepath{fill}%
\end{pgfscope}%
\begin{pgfscope}%
\pgfsetbuttcap%
\pgfsetroundjoin%
\definecolor{currentfill}{rgb}{0.000000,0.000000,0.000000}%
\pgfsetfillcolor{currentfill}%
\pgfsetlinewidth{0.803000pt}%
\definecolor{currentstroke}{rgb}{0.000000,0.000000,0.000000}%
\pgfsetstrokecolor{currentstroke}%
\pgfsetdash{}{0pt}%
\pgfsys@defobject{currentmarker}{\pgfqpoint{0.000000in}{-0.048611in}}{\pgfqpoint{0.000000in}{0.000000in}}{%
\pgfpathmoveto{\pgfqpoint{0.000000in}{0.000000in}}%
\pgfpathlineto{\pgfqpoint{0.000000in}{-0.048611in}}%
\pgfusepath{stroke,fill}%
}%
\begin{pgfscope}%
\pgfsys@transformshift{0.839493in}{0.745593in}%
\pgfsys@useobject{currentmarker}{}%
\end{pgfscope}%
\end{pgfscope}%
\begin{pgfscope}%
\pgftext[x=0.839493in,y=0.648371in,,top]{\rmfamily\fontsize{18.000000}{21.600000}\selectfont 0}%
\end{pgfscope}%
\begin{pgfscope}%
\pgfsetbuttcap%
\pgfsetroundjoin%
\definecolor{currentfill}{rgb}{0.000000,0.000000,0.000000}%
\pgfsetfillcolor{currentfill}%
\pgfsetlinewidth{0.803000pt}%
\definecolor{currentstroke}{rgb}{0.000000,0.000000,0.000000}%
\pgfsetstrokecolor{currentstroke}%
\pgfsetdash{}{0pt}%
\pgfsys@defobject{currentmarker}{\pgfqpoint{0.000000in}{-0.048611in}}{\pgfqpoint{0.000000in}{0.000000in}}{%
\pgfpathmoveto{\pgfqpoint{0.000000in}{0.000000in}}%
\pgfpathlineto{\pgfqpoint{0.000000in}{-0.048611in}}%
\pgfusepath{stroke,fill}%
}%
\begin{pgfscope}%
\pgfsys@transformshift{2.264914in}{0.745593in}%
\pgfsys@useobject{currentmarker}{}%
\end{pgfscope}%
\end{pgfscope}%
\begin{pgfscope}%
\pgftext[x=2.264914in,y=0.648371in,,top]{\rmfamily\fontsize{18.000000}{21.600000}\selectfont Release}%
\end{pgfscope}%
\begin{pgfscope}%
\pgfsetbuttcap%
\pgfsetroundjoin%
\definecolor{currentfill}{rgb}{0.000000,0.000000,0.000000}%
\pgfsetfillcolor{currentfill}%
\pgfsetlinewidth{0.803000pt}%
\definecolor{currentstroke}{rgb}{0.000000,0.000000,0.000000}%
\pgfsetstrokecolor{currentstroke}%
\pgfsetdash{}{0pt}%
\pgfsys@defobject{currentmarker}{\pgfqpoint{0.000000in}{-0.048611in}}{\pgfqpoint{0.000000in}{0.000000in}}{%
\pgfpathmoveto{\pgfqpoint{0.000000in}{0.000000in}}%
\pgfpathlineto{\pgfqpoint{0.000000in}{-0.048611in}}%
\pgfusepath{stroke,fill}%
}%
\begin{pgfscope}%
\pgfsys@transformshift{3.215194in}{0.745593in}%
\pgfsys@useobject{currentmarker}{}%
\end{pgfscope}%
\end{pgfscope}%
\begin{pgfscope}%
\pgftext[x=3.215194in,y=0.648371in,,top]{\rmfamily\fontsize{18.000000}{21.600000}\selectfont Snap}%
\end{pgfscope}%
\begin{pgfscope}%
\pgftext[x=2.027344in,y=0.350872in,,top]{\rmfamily\fontsize{18.000000}{21.600000}\selectfont Time}%
\end{pgfscope}%
\begin{pgfscope}%
\pgfsetbuttcap%
\pgfsetroundjoin%
\definecolor{currentfill}{rgb}{0.000000,0.000000,0.000000}%
\pgfsetfillcolor{currentfill}%
\pgfsetlinewidth{0.803000pt}%
\definecolor{currentstroke}{rgb}{0.000000,0.000000,0.000000}%
\pgfsetstrokecolor{currentstroke}%
\pgfsetdash{}{0pt}%
\pgfsys@defobject{currentmarker}{\pgfqpoint{-0.048611in}{0.000000in}}{\pgfqpoint{0.000000in}{0.000000in}}{%
\pgfpathmoveto{\pgfqpoint{0.000000in}{0.000000in}}%
\pgfpathlineto{\pgfqpoint{-0.048611in}{0.000000in}}%
\pgfusepath{stroke,fill}%
}%
\begin{pgfscope}%
\pgfsys@transformshift{0.407116in}{0.955453in}%
\pgfsys@useobject{currentmarker}{}%
\end{pgfscope}%
\end{pgfscope}%
\begin{pgfscope}%
\pgftext[x=0.272337in,y=0.566571in,left,base,rotate=90.000000]{\rmfamily\fontsize{13.000000}{15.600000}\selectfont Inverted}%
\end{pgfscope}%
\begin{pgfscope}%
\pgfsetbuttcap%
\pgfsetroundjoin%
\definecolor{currentfill}{rgb}{0.000000,0.000000,0.000000}%
\pgfsetfillcolor{currentfill}%
\pgfsetlinewidth{0.803000pt}%
\definecolor{currentstroke}{rgb}{0.000000,0.000000,0.000000}%
\pgfsetstrokecolor{currentstroke}%
\pgfsetdash{}{0pt}%
\pgfsys@defobject{currentmarker}{\pgfqpoint{-0.048611in}{0.000000in}}{\pgfqpoint{0.000000in}{0.000000in}}{%
\pgfpathmoveto{\pgfqpoint{0.000000in}{0.000000in}}%
\pgfpathlineto{\pgfqpoint{-0.048611in}{0.000000in}}%
\pgfusepath{stroke,fill}%
}%
\begin{pgfscope}%
\pgfsys@transformshift{0.407116in}{2.004750in}%
\pgfsys@useobject{currentmarker}{}%
\end{pgfscope}%
\end{pgfscope}%
\begin{pgfscope}%
\pgfsetbuttcap%
\pgfsetroundjoin%
\definecolor{currentfill}{rgb}{0.000000,0.000000,0.000000}%
\pgfsetfillcolor{currentfill}%
\pgfsetlinewidth{0.803000pt}%
\definecolor{currentstroke}{rgb}{0.000000,0.000000,0.000000}%
\pgfsetstrokecolor{currentstroke}%
\pgfsetdash{}{0pt}%
\pgfsys@defobject{currentmarker}{\pgfqpoint{-0.048611in}{0.000000in}}{\pgfqpoint{0.000000in}{0.000000in}}{%
\pgfpathmoveto{\pgfqpoint{0.000000in}{0.000000in}}%
\pgfpathlineto{\pgfqpoint{-0.048611in}{0.000000in}}%
\pgfusepath{stroke,fill}%
}%
\begin{pgfscope}%
\pgfsys@transformshift{0.407116in}{3.054047in}%
\pgfsys@useobject{currentmarker}{}%
\end{pgfscope}%
\end{pgfscope}%
\begin{pgfscope}%
\pgftext[x=0.269780in,y=2.677860in,left,base,rotate=90.000000]{\rmfamily\fontsize{13.000000}{15.600000}\selectfont Straight}%
\end{pgfscope}%
\begin{pgfscope}%
\pgfpathrectangle{\pgfqpoint{0.407116in}{0.745593in}}{\pgfqpoint{3.240456in}{2.518313in}} %
\pgfusepath{clip}%
\pgfsetrectcap%
\pgfsetroundjoin%
\pgfsetlinewidth{0.501875pt}%
\definecolor{currentstroke}{rgb}{0.000000,0.000000,0.000000}%
\pgfsetstrokecolor{currentstroke}%
\pgfsetdash{}{0pt}%
\pgfpathmoveto{\pgfqpoint{0.554409in}{2.004750in}}%
\pgfpathlineto{\pgfqpoint{3.500278in}{2.004750in}}%
\pgfusepath{stroke}%
\end{pgfscope}%
\begin{pgfscope}%
\pgfpathrectangle{\pgfqpoint{0.407116in}{0.745593in}}{\pgfqpoint{3.240456in}{2.518313in}} %
\pgfusepath{clip}%
\pgfsetbuttcap%
\pgfsetroundjoin%
\pgfsetlinewidth{1.003750pt}%
\definecolor{currentstroke}{rgb}{0.000000,0.000000,0.000000}%
\pgfsetstrokecolor{currentstroke}%
\pgfsetstrokeopacity{0.500000}%
\pgfsetdash{{3.700000pt}{1.600000pt}}{0.000000pt}%
\pgfpathmoveto{\pgfqpoint{0.839493in}{0.742260in}}%
\pgfpathlineto{\pgfqpoint{0.839493in}{3.267240in}}%
\pgfusepath{stroke}%
\end{pgfscope}%
\begin{pgfscope}%
\pgfpathrectangle{\pgfqpoint{0.407116in}{0.745593in}}{\pgfqpoint{3.240456in}{2.518313in}} %
\pgfusepath{clip}%
\pgfsetbuttcap%
\pgfsetroundjoin%
\pgfsetlinewidth{1.003750pt}%
\definecolor{currentstroke}{rgb}{0.000000,0.000000,0.000000}%
\pgfsetstrokecolor{currentstroke}%
\pgfsetstrokeopacity{0.500000}%
\pgfsetdash{{3.700000pt}{1.600000pt}}{0.000000pt}%
\pgfpathmoveto{\pgfqpoint{2.264914in}{0.742260in}}%
\pgfpathlineto{\pgfqpoint{2.264914in}{3.267240in}}%
\pgfusepath{stroke}%
\end{pgfscope}%
\begin{pgfscope}%
\pgfpathrectangle{\pgfqpoint{0.407116in}{0.745593in}}{\pgfqpoint{3.240456in}{2.518313in}} %
\pgfusepath{clip}%
\pgfsetbuttcap%
\pgfsetroundjoin%
\pgfsetlinewidth{1.003750pt}%
\definecolor{currentstroke}{rgb}{0.000000,0.000000,0.000000}%
\pgfsetstrokecolor{currentstroke}%
\pgfsetstrokeopacity{0.500000}%
\pgfsetdash{{3.700000pt}{1.600000pt}}{0.000000pt}%
\pgfpathmoveto{\pgfqpoint{3.215194in}{0.742260in}}%
\pgfpathlineto{\pgfqpoint{3.215194in}{3.267240in}}%
\pgfusepath{stroke}%
\end{pgfscope}%
\begin{pgfscope}%
\pgfpathrectangle{\pgfqpoint{0.407116in}{0.745593in}}{\pgfqpoint{3.240456in}{2.518313in}} %
\pgfusepath{clip}%
\pgfsetrectcap%
\pgfsetroundjoin%
\pgfsetlinewidth{2.007500pt}%
\definecolor{currentstroke}{rgb}{0.000000,0.000000,1.000000}%
\pgfsetstrokecolor{currentstroke}%
\pgfsetdash{}{0pt}%
\pgfpathmoveto{\pgfqpoint{0.554409in}{3.054047in}}%
\pgfpathlineto{\pgfqpoint{0.559161in}{3.054047in}}%
\pgfpathlineto{\pgfqpoint{0.563912in}{3.054047in}}%
\pgfpathlineto{\pgfqpoint{0.568663in}{3.054047in}}%
\pgfpathlineto{\pgfqpoint{0.573415in}{3.054047in}}%
\pgfpathlineto{\pgfqpoint{0.578166in}{3.054047in}}%
\pgfpathlineto{\pgfqpoint{0.582918in}{3.054047in}}%
\pgfpathlineto{\pgfqpoint{0.587669in}{3.054047in}}%
\pgfpathlineto{\pgfqpoint{0.592421in}{3.054047in}}%
\pgfpathlineto{\pgfqpoint{0.597172in}{3.054047in}}%
\pgfpathlineto{\pgfqpoint{0.601923in}{3.054047in}}%
\pgfpathlineto{\pgfqpoint{0.606675in}{3.054047in}}%
\pgfpathlineto{\pgfqpoint{0.611426in}{3.054047in}}%
\pgfpathlineto{\pgfqpoint{0.616178in}{3.054047in}}%
\pgfpathlineto{\pgfqpoint{0.620929in}{3.054047in}}%
\pgfpathlineto{\pgfqpoint{0.625680in}{3.054047in}}%
\pgfpathlineto{\pgfqpoint{0.630432in}{3.054047in}}%
\pgfpathlineto{\pgfqpoint{0.635183in}{3.054047in}}%
\pgfpathlineto{\pgfqpoint{0.639935in}{3.054047in}}%
\pgfpathlineto{\pgfqpoint{0.644686in}{3.054047in}}%
\pgfpathlineto{\pgfqpoint{0.649437in}{3.054047in}}%
\pgfpathlineto{\pgfqpoint{0.654189in}{3.054047in}}%
\pgfpathlineto{\pgfqpoint{0.658940in}{3.054047in}}%
\pgfpathlineto{\pgfqpoint{0.663692in}{3.054047in}}%
\pgfpathlineto{\pgfqpoint{0.668443in}{3.054047in}}%
\pgfpathlineto{\pgfqpoint{0.673194in}{3.054047in}}%
\pgfpathlineto{\pgfqpoint{0.677946in}{3.054047in}}%
\pgfpathlineto{\pgfqpoint{0.682697in}{3.054047in}}%
\pgfpathlineto{\pgfqpoint{0.687449in}{3.054047in}}%
\pgfpathlineto{\pgfqpoint{0.692200in}{3.054047in}}%
\pgfpathlineto{\pgfqpoint{0.696951in}{3.054047in}}%
\pgfpathlineto{\pgfqpoint{0.701703in}{3.054047in}}%
\pgfpathlineto{\pgfqpoint{0.706454in}{3.054047in}}%
\pgfpathlineto{\pgfqpoint{0.711206in}{3.054047in}}%
\pgfpathlineto{\pgfqpoint{0.715957in}{3.054047in}}%
\pgfpathlineto{\pgfqpoint{0.720708in}{3.054047in}}%
\pgfpathlineto{\pgfqpoint{0.725460in}{3.054047in}}%
\pgfpathlineto{\pgfqpoint{0.730211in}{3.054047in}}%
\pgfpathlineto{\pgfqpoint{0.734963in}{3.054047in}}%
\pgfpathlineto{\pgfqpoint{0.739714in}{3.054047in}}%
\pgfpathlineto{\pgfqpoint{0.744465in}{3.054047in}}%
\pgfpathlineto{\pgfqpoint{0.749217in}{3.054047in}}%
\pgfpathlineto{\pgfqpoint{0.753968in}{3.054047in}}%
\pgfpathlineto{\pgfqpoint{0.758720in}{3.054047in}}%
\pgfpathlineto{\pgfqpoint{0.763471in}{3.054047in}}%
\pgfpathlineto{\pgfqpoint{0.768222in}{3.054047in}}%
\pgfpathlineto{\pgfqpoint{0.772974in}{3.054047in}}%
\pgfpathlineto{\pgfqpoint{0.777725in}{3.054047in}}%
\pgfpathlineto{\pgfqpoint{0.782477in}{3.054047in}}%
\pgfpathlineto{\pgfqpoint{0.787228in}{3.054047in}}%
\pgfpathlineto{\pgfqpoint{0.791979in}{3.054047in}}%
\pgfpathlineto{\pgfqpoint{0.796731in}{3.054047in}}%
\pgfpathlineto{\pgfqpoint{0.801482in}{3.054047in}}%
\pgfpathlineto{\pgfqpoint{0.806234in}{3.054047in}}%
\pgfpathlineto{\pgfqpoint{0.810985in}{3.054047in}}%
\pgfpathlineto{\pgfqpoint{0.815736in}{3.054047in}}%
\pgfpathlineto{\pgfqpoint{0.820488in}{3.054047in}}%
\pgfpathlineto{\pgfqpoint{0.825239in}{3.054047in}}%
\pgfpathlineto{\pgfqpoint{0.829991in}{3.054047in}}%
\pgfpathlineto{\pgfqpoint{0.834742in}{3.054047in}}%
\pgfusepath{stroke}%
\end{pgfscope}%
\begin{pgfscope}%
\pgfpathrectangle{\pgfqpoint{0.407116in}{0.745593in}}{\pgfqpoint{3.240456in}{2.518313in}} %
\pgfusepath{clip}%
\pgfsetrectcap%
\pgfsetroundjoin%
\pgfsetlinewidth{2.007500pt}%
\definecolor{currentstroke}{rgb}{0.000000,0.000000,1.000000}%
\pgfsetstrokecolor{currentstroke}%
\pgfsetdash{}{0pt}%
\pgfpathmoveto{\pgfqpoint{0.839493in}{0.955453in}}%
\pgfpathlineto{\pgfqpoint{2.260163in}{0.955453in}}%
\pgfpathlineto{\pgfqpoint{2.260163in}{0.955453in}}%
\pgfusepath{stroke}%
\end{pgfscope}%
\begin{pgfscope}%
\pgfpathrectangle{\pgfqpoint{0.407116in}{0.745593in}}{\pgfqpoint{3.240456in}{2.518313in}} %
\pgfusepath{clip}%
\pgfsetrectcap%
\pgfsetroundjoin%
\pgfsetlinewidth{2.007500pt}%
\definecolor{currentstroke}{rgb}{0.000000,0.000000,1.000000}%
\pgfsetstrokecolor{currentstroke}%
\pgfsetdash{}{0pt}%
\pgfpathmoveto{\pgfqpoint{2.264914in}{0.955453in}}%
\pgfpathlineto{\pgfqpoint{2.269665in}{0.986932in}}%
\pgfpathlineto{\pgfqpoint{2.274417in}{0.999971in}}%
\pgfpathlineto{\pgfqpoint{2.279168in}{1.009976in}}%
\pgfpathlineto{\pgfqpoint{2.283920in}{1.018411in}}%
\pgfpathlineto{\pgfqpoint{2.288671in}{1.025842in}}%
\pgfpathlineto{\pgfqpoint{2.298174in}{1.038738in}}%
\pgfpathlineto{\pgfqpoint{2.307677in}{1.049890in}}%
\pgfpathlineto{\pgfqpoint{2.317179in}{1.059857in}}%
\pgfpathlineto{\pgfqpoint{2.326682in}{1.068952in}}%
\pgfpathlineto{\pgfqpoint{2.340936in}{1.081369in}}%
\pgfpathlineto{\pgfqpoint{2.355191in}{1.092666in}}%
\pgfpathlineto{\pgfqpoint{2.369445in}{1.103102in}}%
\pgfpathlineto{\pgfqpoint{2.388450in}{1.115964in}}%
\pgfpathlineto{\pgfqpoint{2.407456in}{1.127870in}}%
\pgfpathlineto{\pgfqpoint{2.426462in}{1.139005in}}%
\pgfpathlineto{\pgfqpoint{2.450219in}{1.152039in}}%
\pgfpathlineto{\pgfqpoint{2.473976in}{1.164260in}}%
\pgfpathlineto{\pgfqpoint{2.502484in}{1.178042in}}%
\pgfpathlineto{\pgfqpoint{2.530992in}{1.191020in}}%
\pgfpathlineto{\pgfqpoint{2.564252in}{1.205309in}}%
\pgfpathlineto{\pgfqpoint{2.597512in}{1.218824in}}%
\pgfpathlineto{\pgfqpoint{2.635523in}{1.233467in}}%
\pgfpathlineto{\pgfqpoint{2.678286in}{1.249069in}}%
\pgfpathlineto{\pgfqpoint{2.721049in}{1.263882in}}%
\pgfpathlineto{\pgfqpoint{2.768563in}{1.279548in}}%
\pgfpathlineto{\pgfqpoint{2.820828in}{1.295949in}}%
\pgfpathlineto{\pgfqpoint{2.873093in}{1.311596in}}%
\pgfpathlineto{\pgfqpoint{2.930110in}{1.327916in}}%
\pgfpathlineto{\pgfqpoint{2.991878in}{1.344826in}}%
\pgfpathlineto{\pgfqpoint{3.058398in}{1.362250in}}%
\pgfpathlineto{\pgfqpoint{3.129669in}{1.380127in}}%
\pgfpathlineto{\pgfqpoint{3.205692in}{1.398401in}}%
\pgfpathlineto{\pgfqpoint{3.210443in}{1.399518in}}%
\pgfpathlineto{\pgfqpoint{3.210443in}{1.399518in}}%
\pgfusepath{stroke}%
\end{pgfscope}%
\begin{pgfscope}%
\pgfpathrectangle{\pgfqpoint{0.407116in}{0.745593in}}{\pgfqpoint{3.240456in}{2.518313in}} %
\pgfusepath{clip}%
\pgfsetrectcap%
\pgfsetroundjoin%
\pgfsetlinewidth{2.007500pt}%
\definecolor{currentstroke}{rgb}{0.000000,0.000000,1.000000}%
\pgfsetstrokecolor{currentstroke}%
\pgfsetdash{}{0pt}%
\pgfpathmoveto{\pgfqpoint{3.215194in}{3.054047in}}%
\pgfpathlineto{\pgfqpoint{3.219946in}{3.054047in}}%
\pgfpathlineto{\pgfqpoint{3.224697in}{3.054047in}}%
\pgfpathlineto{\pgfqpoint{3.229449in}{3.054047in}}%
\pgfpathlineto{\pgfqpoint{3.234200in}{3.054047in}}%
\pgfpathlineto{\pgfqpoint{3.238951in}{3.054047in}}%
\pgfpathlineto{\pgfqpoint{3.243703in}{3.054047in}}%
\pgfpathlineto{\pgfqpoint{3.248454in}{3.054047in}}%
\pgfpathlineto{\pgfqpoint{3.253206in}{3.054047in}}%
\pgfpathlineto{\pgfqpoint{3.257957in}{3.054047in}}%
\pgfpathlineto{\pgfqpoint{3.262708in}{3.054047in}}%
\pgfpathlineto{\pgfqpoint{3.267460in}{3.054047in}}%
\pgfpathlineto{\pgfqpoint{3.272211in}{3.054047in}}%
\pgfpathlineto{\pgfqpoint{3.276963in}{3.054047in}}%
\pgfpathlineto{\pgfqpoint{3.281714in}{3.054047in}}%
\pgfpathlineto{\pgfqpoint{3.286465in}{3.054047in}}%
\pgfpathlineto{\pgfqpoint{3.291217in}{3.054047in}}%
\pgfpathlineto{\pgfqpoint{3.295968in}{3.054047in}}%
\pgfpathlineto{\pgfqpoint{3.300720in}{3.054047in}}%
\pgfpathlineto{\pgfqpoint{3.305471in}{3.054047in}}%
\pgfpathlineto{\pgfqpoint{3.310222in}{3.054047in}}%
\pgfpathlineto{\pgfqpoint{3.314974in}{3.054047in}}%
\pgfpathlineto{\pgfqpoint{3.319725in}{3.054047in}}%
\pgfpathlineto{\pgfqpoint{3.324477in}{3.054047in}}%
\pgfpathlineto{\pgfqpoint{3.329228in}{3.054047in}}%
\pgfpathlineto{\pgfqpoint{3.333979in}{3.054047in}}%
\pgfpathlineto{\pgfqpoint{3.338731in}{3.054047in}}%
\pgfpathlineto{\pgfqpoint{3.343482in}{3.054047in}}%
\pgfpathlineto{\pgfqpoint{3.348234in}{3.054047in}}%
\pgfpathlineto{\pgfqpoint{3.352985in}{3.054047in}}%
\pgfpathlineto{\pgfqpoint{3.357736in}{3.054047in}}%
\pgfpathlineto{\pgfqpoint{3.362488in}{3.054047in}}%
\pgfpathlineto{\pgfqpoint{3.367239in}{3.054047in}}%
\pgfpathlineto{\pgfqpoint{3.371991in}{3.054047in}}%
\pgfpathlineto{\pgfqpoint{3.376742in}{3.054047in}}%
\pgfpathlineto{\pgfqpoint{3.381493in}{3.054047in}}%
\pgfpathlineto{\pgfqpoint{3.386245in}{3.054047in}}%
\pgfpathlineto{\pgfqpoint{3.390996in}{3.054047in}}%
\pgfpathlineto{\pgfqpoint{3.395748in}{3.054047in}}%
\pgfpathlineto{\pgfqpoint{3.400499in}{3.054047in}}%
\pgfpathlineto{\pgfqpoint{3.405250in}{3.054047in}}%
\pgfpathlineto{\pgfqpoint{3.410002in}{3.054047in}}%
\pgfpathlineto{\pgfqpoint{3.414753in}{3.054047in}}%
\pgfpathlineto{\pgfqpoint{3.419505in}{3.054047in}}%
\pgfpathlineto{\pgfqpoint{3.424256in}{3.054047in}}%
\pgfpathlineto{\pgfqpoint{3.429007in}{3.054047in}}%
\pgfpathlineto{\pgfqpoint{3.433759in}{3.054047in}}%
\pgfpathlineto{\pgfqpoint{3.438510in}{3.054047in}}%
\pgfpathlineto{\pgfqpoint{3.443262in}{3.054047in}}%
\pgfpathlineto{\pgfqpoint{3.448013in}{3.054047in}}%
\pgfpathlineto{\pgfqpoint{3.452764in}{3.054047in}}%
\pgfpathlineto{\pgfqpoint{3.457516in}{3.054047in}}%
\pgfpathlineto{\pgfqpoint{3.462267in}{3.054047in}}%
\pgfpathlineto{\pgfqpoint{3.467019in}{3.054047in}}%
\pgfpathlineto{\pgfqpoint{3.471770in}{3.054047in}}%
\pgfpathlineto{\pgfqpoint{3.476521in}{3.054047in}}%
\pgfpathlineto{\pgfqpoint{3.481273in}{3.054047in}}%
\pgfpathlineto{\pgfqpoint{3.486024in}{3.054047in}}%
\pgfpathlineto{\pgfqpoint{3.490776in}{3.054047in}}%
\pgfpathlineto{\pgfqpoint{3.495527in}{3.054047in}}%
\pgfusepath{stroke}%
\end{pgfscope}%
\begin{pgfscope}%
\pgfsetrectcap%
\pgfsetmiterjoin%
\pgfsetlinewidth{0.803000pt}%
\definecolor{currentstroke}{rgb}{0.000000,0.000000,0.000000}%
\pgfsetstrokecolor{currentstroke}%
\pgfsetdash{}{0pt}%
\pgfpathmoveto{\pgfqpoint{0.407116in}{0.745593in}}%
\pgfpathlineto{\pgfqpoint{0.407116in}{3.263907in}}%
\pgfusepath{stroke}%
\end{pgfscope}%
\begin{pgfscope}%
\pgfsetrectcap%
\pgfsetmiterjoin%
\pgfsetlinewidth{0.803000pt}%
\definecolor{currentstroke}{rgb}{0.000000,0.000000,0.000000}%
\pgfsetstrokecolor{currentstroke}%
\pgfsetdash{}{0pt}%
\pgfpathmoveto{\pgfqpoint{3.647572in}{0.745593in}}%
\pgfpathlineto{\pgfqpoint{3.647572in}{3.263907in}}%
\pgfusepath{stroke}%
\end{pgfscope}%
\begin{pgfscope}%
\pgfsetrectcap%
\pgfsetmiterjoin%
\pgfsetlinewidth{0.803000pt}%
\definecolor{currentstroke}{rgb}{0.000000,0.000000,0.000000}%
\pgfsetstrokecolor{currentstroke}%
\pgfsetdash{}{0pt}%
\pgfpathmoveto{\pgfqpoint{0.407116in}{0.745593in}}%
\pgfpathlineto{\pgfqpoint{3.647572in}{0.745593in}}%
\pgfusepath{stroke}%
\end{pgfscope}%
\begin{pgfscope}%
\pgfsetrectcap%
\pgfsetmiterjoin%
\pgfsetlinewidth{0.803000pt}%
\definecolor{currentstroke}{rgb}{0.000000,0.000000,0.000000}%
\pgfsetstrokecolor{currentstroke}%
\pgfsetdash{}{0pt}%
\pgfpathmoveto{\pgfqpoint{0.407116in}{3.263907in}}%
\pgfpathlineto{\pgfqpoint{3.647572in}{3.263907in}}%
\pgfusepath{stroke}%
\end{pgfscope}%
\begin{pgfscope}%
\pgftext[x=1.552204in,y=2.319539in,,base]{\rmfamily\fontsize{20.000000}{24.000000}\selectfont \(\displaystyle t_\mathrm{hold}\)}%
\end{pgfscope}%
\begin{pgfscope}%
\pgftext[x=2.740054in,y=2.319539in,,base]{\rmfamily\fontsize{20.000000}{24.000000}\selectfont \(\displaystyle t_\mathrm{flip}\)}%
\end{pgfscope}%
\end{pgfpicture}%
\makeatother%
\endgroup%

%% file: flip_time_vs_thickness.pgf
\begingroup%
\makeatletter%
\begin{pgfpicture}%
\pgfpathrectangle{\pgfpointorigin}{\pgfqpoint{6.000000in}{3.000000in}}%
\pgfusepath{use as bounding box, clip}%
\begin{pgfscope}%
\pgfsetbuttcap%
\pgfsetmiterjoin%
\definecolor{currentfill}{rgb}{1.000000,1.000000,1.000000}%
\pgfsetfillcolor{currentfill}%
\pgfsetlinewidth{0.000000pt}%
\definecolor{currentstroke}{rgb}{1.000000,1.000000,1.000000}%
\pgfsetstrokecolor{currentstroke}%
\pgfsetdash{}{0pt}%
\pgfpathmoveto{\pgfqpoint{0.000000in}{0.000000in}}%
\pgfpathlineto{\pgfqpoint{6.000000in}{0.000000in}}%
\pgfpathlineto{\pgfqpoint{6.000000in}{3.000000in}}%
\pgfpathlineto{\pgfqpoint{0.000000in}{3.000000in}}%
\pgfpathclose%
\pgfusepath{fill}%
\end{pgfscope}%
\begin{pgfscope}%
\pgfsetbuttcap%
\pgfsetmiterjoin%
\definecolor{currentfill}{rgb}{1.000000,1.000000,1.000000}%
\pgfsetfillcolor{currentfill}%
\pgfsetlinewidth{0.000000pt}%
\definecolor{currentstroke}{rgb}{0.000000,0.000000,0.000000}%
\pgfsetstrokecolor{currentstroke}%
\pgfsetstrokeopacity{0.000000}%
\pgfsetdash{}{0pt}%
\pgfpathmoveto{\pgfqpoint{0.842041in}{0.773252in}}%
\pgfpathlineto{\pgfqpoint{5.815000in}{0.773252in}}%
\pgfpathlineto{\pgfqpoint{5.815000in}{2.763519in}}%
\pgfpathlineto{\pgfqpoint{0.842041in}{2.763519in}}%
\pgfpathclose%
\pgfusepath{fill}%
\end{pgfscope}%
\begin{pgfscope}%
\pgfpathrectangle{\pgfqpoint{0.842041in}{0.773252in}}{\pgfqpoint{4.972959in}{1.990267in}} %
\pgfusepath{clip}%
\pgfsetbuttcap%
\pgfsetmiterjoin%
\definecolor{currentfill}{rgb}{0.752900,0.849010,0.958700}%
\pgfsetfillcolor{currentfill}%
\pgfsetlinewidth{1.003750pt}%
\definecolor{currentstroke}{rgb}{0.752900,0.849010,0.958700}%
\pgfsetstrokecolor{currentstroke}%
\pgfsetdash{}{0pt}%
\pgfpathmoveto{\pgfqpoint{2.386093in}{-5.197550in}}%
\pgfpathlineto{\pgfqpoint{2.386093in}{8.887418in}}%
\pgfpathlineto{\pgfqpoint{4.383769in}{8.887418in}}%
\pgfpathlineto{\pgfqpoint{4.383769in}{-5.197550in}}%
\pgfpathclose%
\pgfusepath{stroke,fill}%
\end{pgfscope}%
\begin{pgfscope}%
\pgfpathrectangle{\pgfqpoint{0.842041in}{0.773252in}}{\pgfqpoint{4.972959in}{1.990267in}} %
\pgfusepath{clip}%
\pgfsetbuttcap%
\pgfsetmiterjoin%
\definecolor{currentfill}{rgb}{1.000000,0.900000,0.900000}%
\pgfsetfillcolor{currentfill}%
\pgfsetlinewidth{1.003750pt}%
\definecolor{currentstroke}{rgb}{1.000000,0.900000,0.900000}%
\pgfsetstrokecolor{currentstroke}%
\pgfsetdash{}{0pt}%
\pgfpathmoveto{\pgfqpoint{4.383769in}{-5.197550in}}%
\pgfpathlineto{\pgfqpoint{4.383769in}{8.887418in}}%
\pgfpathlineto{\pgfqpoint{96.232433in}{8.887418in}}%
\pgfpathlineto{\pgfqpoint{96.232433in}{-5.197550in}}%
\pgfpathclose%
\pgfusepath{stroke,fill}%
\end{pgfscope}%
\begin{pgfscope}%
\pgfsetbuttcap%
\pgfsetroundjoin%
\definecolor{currentfill}{rgb}{0.000000,0.000000,0.000000}%
\pgfsetfillcolor{currentfill}%
\pgfsetlinewidth{0.803000pt}%
\definecolor{currentstroke}{rgb}{0.000000,0.000000,0.000000}%
\pgfsetstrokecolor{currentstroke}%
\pgfsetdash{}{0pt}%
\pgfsys@defobject{currentmarker}{\pgfqpoint{0.000000in}{-0.048611in}}{\pgfqpoint{0.000000in}{0.000000in}}{%
\pgfpathmoveto{\pgfqpoint{0.000000in}{0.000000in}}%
\pgfpathlineto{\pgfqpoint{0.000000in}{-0.048611in}}%
\pgfusepath{stroke,fill}%
}%
\begin{pgfscope}%
\pgfsys@transformshift{0.842041in}{0.773252in}%
\pgfsys@useobject{currentmarker}{}%
\end{pgfscope}%
\end{pgfscope}%
\begin{pgfscope}%
\pgftext[x=0.842041in,y=0.676029in,,top]{\rmfamily\fontsize{18.000000}{21.600000}\selectfont \(\displaystyle 0.50\)}%
\end{pgfscope}%
\begin{pgfscope}%
\pgfsetbuttcap%
\pgfsetroundjoin%
\definecolor{currentfill}{rgb}{0.000000,0.000000,0.000000}%
\pgfsetfillcolor{currentfill}%
\pgfsetlinewidth{0.803000pt}%
\definecolor{currentstroke}{rgb}{0.000000,0.000000,0.000000}%
\pgfsetstrokecolor{currentstroke}%
\pgfsetdash{}{0pt}%
\pgfsys@defobject{currentmarker}{\pgfqpoint{0.000000in}{-0.048611in}}{\pgfqpoint{0.000000in}{0.000000in}}{%
\pgfpathmoveto{\pgfqpoint{0.000000in}{0.000000in}}%
\pgfpathlineto{\pgfqpoint{0.000000in}{-0.048611in}}%
\pgfusepath{stroke,fill}%
}%
\begin{pgfscope}%
\pgfsys@transformshift{1.746215in}{0.773252in}%
\pgfsys@useobject{currentmarker}{}%
\end{pgfscope}%
\end{pgfscope}%
\begin{pgfscope}%
\pgftext[x=1.746215in,y=0.676029in,,top]{\rmfamily\fontsize{18.000000}{21.600000}\selectfont \(\displaystyle 0.52\)}%
\end{pgfscope}%
\begin{pgfscope}%
\pgfsetbuttcap%
\pgfsetroundjoin%
\definecolor{currentfill}{rgb}{0.000000,0.000000,0.000000}%
\pgfsetfillcolor{currentfill}%
\pgfsetlinewidth{0.803000pt}%
\definecolor{currentstroke}{rgb}{0.000000,0.000000,0.000000}%
\pgfsetstrokecolor{currentstroke}%
\pgfsetdash{}{0pt}%
\pgfsys@defobject{currentmarker}{\pgfqpoint{0.000000in}{-0.048611in}}{\pgfqpoint{0.000000in}{0.000000in}}{%
\pgfpathmoveto{\pgfqpoint{0.000000in}{0.000000in}}%
\pgfpathlineto{\pgfqpoint{0.000000in}{-0.048611in}}%
\pgfusepath{stroke,fill}%
}%
\begin{pgfscope}%
\pgfsys@transformshift{2.650390in}{0.773252in}%
\pgfsys@useobject{currentmarker}{}%
\end{pgfscope}%
\end{pgfscope}%
\begin{pgfscope}%
\pgftext[x=2.650390in,y=0.676029in,,top]{\rmfamily\fontsize{18.000000}{21.600000}\selectfont \(\displaystyle 0.54\)}%
\end{pgfscope}%
\begin{pgfscope}%
\pgfsetbuttcap%
\pgfsetroundjoin%
\definecolor{currentfill}{rgb}{0.000000,0.000000,0.000000}%
\pgfsetfillcolor{currentfill}%
\pgfsetlinewidth{0.803000pt}%
\definecolor{currentstroke}{rgb}{0.000000,0.000000,0.000000}%
\pgfsetstrokecolor{currentstroke}%
\pgfsetdash{}{0pt}%
\pgfsys@defobject{currentmarker}{\pgfqpoint{0.000000in}{-0.048611in}}{\pgfqpoint{0.000000in}{0.000000in}}{%
\pgfpathmoveto{\pgfqpoint{0.000000in}{0.000000in}}%
\pgfpathlineto{\pgfqpoint{0.000000in}{-0.048611in}}%
\pgfusepath{stroke,fill}%
}%
\begin{pgfscope}%
\pgfsys@transformshift{3.554564in}{0.773252in}%
\pgfsys@useobject{currentmarker}{}%
\end{pgfscope}%
\end{pgfscope}%
\begin{pgfscope}%
\pgftext[x=3.554564in,y=0.676029in,,top]{\rmfamily\fontsize{18.000000}{21.600000}\selectfont \(\displaystyle 0.56\)}%
\end{pgfscope}%
\begin{pgfscope}%
\pgfsetbuttcap%
\pgfsetroundjoin%
\definecolor{currentfill}{rgb}{0.000000,0.000000,0.000000}%
\pgfsetfillcolor{currentfill}%
\pgfsetlinewidth{0.803000pt}%
\definecolor{currentstroke}{rgb}{0.000000,0.000000,0.000000}%
\pgfsetstrokecolor{currentstroke}%
\pgfsetdash{}{0pt}%
\pgfsys@defobject{currentmarker}{\pgfqpoint{0.000000in}{-0.048611in}}{\pgfqpoint{0.000000in}{0.000000in}}{%
\pgfpathmoveto{\pgfqpoint{0.000000in}{0.000000in}}%
\pgfpathlineto{\pgfqpoint{0.000000in}{-0.048611in}}%
\pgfusepath{stroke,fill}%
}%
\begin{pgfscope}%
\pgfsys@transformshift{4.458738in}{0.773252in}%
\pgfsys@useobject{currentmarker}{}%
\end{pgfscope}%
\end{pgfscope}%
\begin{pgfscope}%
\pgftext[x=4.458738in,y=0.676029in,,top]{\rmfamily\fontsize{18.000000}{21.600000}\selectfont \(\displaystyle 0.58\)}%
\end{pgfscope}%
\begin{pgfscope}%
\pgfsetbuttcap%
\pgfsetroundjoin%
\definecolor{currentfill}{rgb}{0.000000,0.000000,0.000000}%
\pgfsetfillcolor{currentfill}%
\pgfsetlinewidth{0.803000pt}%
\definecolor{currentstroke}{rgb}{0.000000,0.000000,0.000000}%
\pgfsetstrokecolor{currentstroke}%
\pgfsetdash{}{0pt}%
\pgfsys@defobject{currentmarker}{\pgfqpoint{0.000000in}{-0.048611in}}{\pgfqpoint{0.000000in}{0.000000in}}{%
\pgfpathmoveto{\pgfqpoint{0.000000in}{0.000000in}}%
\pgfpathlineto{\pgfqpoint{0.000000in}{-0.048611in}}%
\pgfusepath{stroke,fill}%
}%
\begin{pgfscope}%
\pgfsys@transformshift{5.362913in}{0.773252in}%
\pgfsys@useobject{currentmarker}{}%
\end{pgfscope}%
\end{pgfscope}%
\begin{pgfscope}%
\pgftext[x=5.362913in,y=0.676029in,,top]{\rmfamily\fontsize{18.000000}{21.600000}\selectfont \(\displaystyle 0.60\)}%
\end{pgfscope}%
\begin{pgfscope}%
\pgftext[x=3.328521in,y=0.378531in,,top]{\rmfamily\fontsize{18.000000}{21.600000}\selectfont \(\displaystyle h/r_\mathrm{min}\)}%
\end{pgfscope}%
\begin{pgfscope}%
\pgfsetbuttcap%
\pgfsetroundjoin%
\definecolor{currentfill}{rgb}{0.000000,0.000000,0.000000}%
\pgfsetfillcolor{currentfill}%
\pgfsetlinewidth{0.803000pt}%
\definecolor{currentstroke}{rgb}{0.000000,0.000000,0.000000}%
\pgfsetstrokecolor{currentstroke}%
\pgfsetdash{}{0pt}%
\pgfsys@defobject{currentmarker}{\pgfqpoint{-0.048611in}{0.000000in}}{\pgfqpoint{0.000000in}{0.000000in}}{%
\pgfpathmoveto{\pgfqpoint{0.000000in}{0.000000in}}%
\pgfpathlineto{\pgfqpoint{-0.048611in}{0.000000in}}%
\pgfusepath{stroke,fill}%
}%
\begin{pgfscope}%
\pgfsys@transformshift{0.842041in}{0.926349in}%
\pgfsys@useobject{currentmarker}{}%
\end{pgfscope}%
\end{pgfscope}%
\begin{pgfscope}%
\pgftext[x=0.347236in,y=0.831378in,left,base]{\rmfamily\fontsize{18.000000}{21.600000}\selectfont 0.0}%
\end{pgfscope}%
\begin{pgfscope}%
\pgfsetbuttcap%
\pgfsetroundjoin%
\definecolor{currentfill}{rgb}{0.000000,0.000000,0.000000}%
\pgfsetfillcolor{currentfill}%
\pgfsetlinewidth{0.803000pt}%
\definecolor{currentstroke}{rgb}{0.000000,0.000000,0.000000}%
\pgfsetstrokecolor{currentstroke}%
\pgfsetdash{}{0pt}%
\pgfsys@defobject{currentmarker}{\pgfqpoint{-0.048611in}{0.000000in}}{\pgfqpoint{0.000000in}{0.000000in}}{%
\pgfpathmoveto{\pgfqpoint{0.000000in}{0.000000in}}%
\pgfpathlineto{\pgfqpoint{-0.048611in}{0.000000in}}%
\pgfusepath{stroke,fill}%
}%
\begin{pgfscope}%
\pgfsys@transformshift{0.842041in}{1.538739in}%
\pgfsys@useobject{currentmarker}{}%
\end{pgfscope}%
\end{pgfscope}%
\begin{pgfscope}%
\pgftext[x=0.347236in,y=1.443768in,left,base]{\rmfamily\fontsize{18.000000}{21.600000}\selectfont 0.2}%
\end{pgfscope}%
\begin{pgfscope}%
\pgfsetbuttcap%
\pgfsetroundjoin%
\definecolor{currentfill}{rgb}{0.000000,0.000000,0.000000}%
\pgfsetfillcolor{currentfill}%
\pgfsetlinewidth{0.803000pt}%
\definecolor{currentstroke}{rgb}{0.000000,0.000000,0.000000}%
\pgfsetstrokecolor{currentstroke}%
\pgfsetdash{}{0pt}%
\pgfsys@defobject{currentmarker}{\pgfqpoint{-0.048611in}{0.000000in}}{\pgfqpoint{0.000000in}{0.000000in}}{%
\pgfpathmoveto{\pgfqpoint{0.000000in}{0.000000in}}%
\pgfpathlineto{\pgfqpoint{-0.048611in}{0.000000in}}%
\pgfusepath{stroke,fill}%
}%
\begin{pgfscope}%
\pgfsys@transformshift{0.842041in}{2.151129in}%
\pgfsys@useobject{currentmarker}{}%
\end{pgfscope}%
\end{pgfscope}%
\begin{pgfscope}%
\pgftext[x=0.347236in,y=2.056158in,left,base]{\rmfamily\fontsize{18.000000}{21.600000}\selectfont 0.4}%
\end{pgfscope}%
\begin{pgfscope}%
\pgfsetbuttcap%
\pgfsetroundjoin%
\definecolor{currentfill}{rgb}{0.000000,0.000000,0.000000}%
\pgfsetfillcolor{currentfill}%
\pgfsetlinewidth{0.803000pt}%
\definecolor{currentstroke}{rgb}{0.000000,0.000000,0.000000}%
\pgfsetstrokecolor{currentstroke}%
\pgfsetdash{}{0pt}%
\pgfsys@defobject{currentmarker}{\pgfqpoint{-0.048611in}{0.000000in}}{\pgfqpoint{0.000000in}{0.000000in}}{%
\pgfpathmoveto{\pgfqpoint{0.000000in}{0.000000in}}%
\pgfpathlineto{\pgfqpoint{-0.048611in}{0.000000in}}%
\pgfusepath{stroke,fill}%
}%
\begin{pgfscope}%
\pgfsys@transformshift{0.842041in}{2.763519in}%
\pgfsys@useobject{currentmarker}{}%
\end{pgfscope}%
\end{pgfscope}%
\begin{pgfscope}%
\pgftext[x=0.504818in,y=2.668548in,left,base]{\rmfamily\fontsize{18.000000}{21.600000}\selectfont \(\displaystyle \infty\)}%
\end{pgfscope}%
\begin{pgfscope}%
\pgftext[x=0.291680in,y=1.768385in,,bottom,rotate=90.000000]{\rmfamily\fontsize{18.000000}{21.600000}\selectfont Flip time [sec]}%
\end{pgfscope}%
\begin{pgfscope}%
\pgfpathrectangle{\pgfqpoint{0.842041in}{0.773252in}}{\pgfqpoint{4.972959in}{1.990267in}} %
\pgfusepath{clip}%
\pgfsetbuttcap%
\pgfsetroundjoin%
\pgfsetlinewidth{2.007500pt}%
\definecolor{currentstroke}{rgb}{0.000000,0.000000,0.000000}%
\pgfsetstrokecolor{currentstroke}%
\pgfsetdash{{2.000000pt}{3.300000pt}}{0.000000pt}%
\pgfpathmoveto{\pgfqpoint{2.474272in}{2.766852in}}%
\pgfpathlineto{\pgfqpoint{2.475791in}{2.226524in}}%
\pgfusepath{stroke}%
\end{pgfscope}%
\begin{pgfscope}%
\pgfpathrectangle{\pgfqpoint{0.842041in}{0.773252in}}{\pgfqpoint{4.972959in}{1.990267in}} %
\pgfusepath{clip}%
\pgfsetrectcap%
\pgfsetroundjoin%
\pgfsetlinewidth{2.007500pt}%
\definecolor{currentstroke}{rgb}{0.000000,0.000000,0.000000}%
\pgfsetstrokecolor{currentstroke}%
\pgfsetdash{}{0pt}%
\pgfpathmoveto{\pgfqpoint{2.469555in}{2.249604in}}%
\pgfpathlineto{\pgfqpoint{2.475791in}{2.226524in}}%
\pgfpathlineto{\pgfqpoint{2.482026in}{2.195750in}}%
\pgfpathlineto{\pgfqpoint{2.488262in}{2.172670in}}%
\pgfpathlineto{\pgfqpoint{2.494498in}{2.134204in}}%
\pgfpathlineto{\pgfqpoint{2.500733in}{2.134204in}}%
\pgfpathlineto{\pgfqpoint{2.506969in}{2.103430in}}%
\pgfpathlineto{\pgfqpoint{2.513205in}{2.064963in}}%
\pgfpathlineto{\pgfqpoint{2.519440in}{2.041883in}}%
\pgfpathlineto{\pgfqpoint{2.525676in}{2.026497in}}%
\pgfpathlineto{\pgfqpoint{2.531912in}{2.018803in}}%
\pgfpathlineto{\pgfqpoint{2.544383in}{1.972643in}}%
\pgfpathlineto{\pgfqpoint{2.550619in}{1.957257in}}%
\pgfpathlineto{\pgfqpoint{2.556855in}{1.949563in}}%
\pgfpathlineto{\pgfqpoint{2.563090in}{1.918790in}}%
\pgfpathlineto{\pgfqpoint{2.569326in}{1.911097in}}%
\pgfpathlineto{\pgfqpoint{2.575562in}{1.895710in}}%
\pgfpathlineto{\pgfqpoint{2.581797in}{1.872630in}}%
\pgfpathlineto{\pgfqpoint{2.594269in}{1.841857in}}%
\pgfpathlineto{\pgfqpoint{2.600504in}{1.849550in}}%
\pgfpathlineto{\pgfqpoint{2.606740in}{1.818777in}}%
\pgfpathlineto{\pgfqpoint{2.612976in}{1.803390in}}%
\pgfpathlineto{\pgfqpoint{2.631683in}{1.780310in}}%
\pgfpathlineto{\pgfqpoint{2.637918in}{1.757230in}}%
\pgfpathlineto{\pgfqpoint{2.650390in}{1.757230in}}%
\pgfpathlineto{\pgfqpoint{2.666292in}{1.715033in}}%
\pgfpathlineto{\pgfqpoint{2.698098in}{1.659361in}}%
\pgfpathlineto{\pgfqpoint{2.825318in}{1.510903in}}%
\pgfpathlineto{\pgfqpoint{2.841221in}{1.501624in}}%
\pgfpathlineto{\pgfqpoint{2.857123in}{1.483067in}}%
\pgfpathlineto{\pgfqpoint{2.873026in}{1.473789in}}%
\pgfpathlineto{\pgfqpoint{2.888928in}{1.455231in}}%
\pgfpathlineto{\pgfqpoint{2.904831in}{1.445953in}}%
\pgfpathlineto{\pgfqpoint{2.920733in}{1.427395in}}%
\pgfpathlineto{\pgfqpoint{2.936636in}{1.418117in}}%
\pgfpathlineto{\pgfqpoint{2.952539in}{1.418117in}}%
\pgfpathlineto{\pgfqpoint{2.968441in}{1.390281in}}%
\pgfpathlineto{\pgfqpoint{3.016149in}{1.362445in}}%
\pgfpathlineto{\pgfqpoint{3.032051in}{1.362445in}}%
\pgfpathlineto{\pgfqpoint{3.047954in}{1.343888in}}%
\pgfpathlineto{\pgfqpoint{3.127467in}{1.297494in}}%
\pgfpathlineto{\pgfqpoint{3.143369in}{1.297494in}}%
\pgfpathlineto{\pgfqpoint{3.159272in}{1.278937in}}%
\pgfpathlineto{\pgfqpoint{3.175174in}{1.278937in}}%
\pgfpathlineto{\pgfqpoint{3.191077in}{1.269659in}}%
\pgfpathlineto{\pgfqpoint{3.206980in}{1.251101in}}%
\pgfpathlineto{\pgfqpoint{3.222882in}{1.251101in}}%
\pgfpathlineto{\pgfqpoint{3.286492in}{1.213987in}}%
\pgfpathlineto{\pgfqpoint{3.302395in}{1.213987in}}%
\pgfpathlineto{\pgfqpoint{3.350103in}{1.186151in}}%
\pgfpathlineto{\pgfqpoint{3.366005in}{1.186151in}}%
\pgfpathlineto{\pgfqpoint{3.381908in}{1.176872in}}%
\pgfpathlineto{\pgfqpoint{3.397810in}{1.176872in}}%
\pgfpathlineto{\pgfqpoint{3.429615in}{1.158315in}}%
\pgfpathlineto{\pgfqpoint{3.445518in}{1.158315in}}%
\pgfpathlineto{\pgfqpoint{3.477323in}{1.139758in}}%
\pgfpathlineto{\pgfqpoint{3.493226in}{1.139758in}}%
\pgfpathlineto{\pgfqpoint{3.509128in}{1.130479in}}%
\pgfpathlineto{\pgfqpoint{3.525031in}{1.130479in}}%
\pgfpathlineto{\pgfqpoint{3.556836in}{1.111922in}}%
\pgfpathlineto{\pgfqpoint{3.588641in}{1.111922in}}%
\pgfpathlineto{\pgfqpoint{3.620446in}{1.093365in}}%
\pgfpathlineto{\pgfqpoint{3.636349in}{1.093365in}}%
\pgfpathlineto{\pgfqpoint{3.652251in}{1.084086in}}%
\pgfpathlineto{\pgfqpoint{3.668154in}{1.084086in}}%
\pgfpathlineto{\pgfqpoint{3.684056in}{1.074807in}}%
\pgfpathlineto{\pgfqpoint{3.699959in}{1.074807in}}%
\pgfpathlineto{\pgfqpoint{3.715862in}{1.065529in}}%
\pgfpathlineto{\pgfqpoint{3.731764in}{1.065529in}}%
\pgfpathlineto{\pgfqpoint{3.747667in}{1.056250in}}%
\pgfpathlineto{\pgfqpoint{3.779472in}{1.056250in}}%
\pgfpathlineto{\pgfqpoint{3.795374in}{1.046971in}}%
\pgfpathlineto{\pgfqpoint{3.811277in}{1.046971in}}%
\pgfpathlineto{\pgfqpoint{3.827180in}{1.037693in}}%
\pgfpathlineto{\pgfqpoint{3.843082in}{1.037693in}}%
\pgfpathlineto{\pgfqpoint{3.858985in}{1.028414in}}%
\pgfpathlineto{\pgfqpoint{3.890790in}{1.028414in}}%
\pgfpathlineto{\pgfqpoint{3.906692in}{1.019135in}}%
\pgfpathlineto{\pgfqpoint{3.922595in}{1.019135in}}%
\pgfpathlineto{\pgfqpoint{3.938497in}{1.009857in}}%
\pgfpathlineto{\pgfqpoint{3.970303in}{1.009857in}}%
\pgfpathlineto{\pgfqpoint{3.986205in}{1.000578in}}%
\pgfpathlineto{\pgfqpoint{4.018010in}{1.000578in}}%
\pgfpathlineto{\pgfqpoint{4.033913in}{0.991300in}}%
\pgfpathlineto{\pgfqpoint{4.065718in}{0.991300in}}%
\pgfpathlineto{\pgfqpoint{4.081621in}{0.982021in}}%
\pgfpathlineto{\pgfqpoint{4.097523in}{0.982021in}}%
\pgfpathlineto{\pgfqpoint{4.113426in}{0.972742in}}%
\pgfpathlineto{\pgfqpoint{4.161133in}{0.972742in}}%
\pgfpathlineto{\pgfqpoint{4.177036in}{0.963464in}}%
\pgfpathlineto{\pgfqpoint{4.208841in}{0.963464in}}%
\pgfpathlineto{\pgfqpoint{4.224744in}{0.954185in}}%
\pgfpathlineto{\pgfqpoint{4.256549in}{0.954185in}}%
\pgfpathlineto{\pgfqpoint{4.272451in}{0.944906in}}%
\pgfpathlineto{\pgfqpoint{4.320159in}{0.944906in}}%
\pgfpathlineto{\pgfqpoint{4.336062in}{0.935628in}}%
\pgfpathlineto{\pgfqpoint{4.367867in}{0.935628in}}%
\pgfpathlineto{\pgfqpoint{4.383769in}{0.926349in}}%
\pgfpathlineto{\pgfqpoint{5.815000in}{0.926349in}}%
\pgfpathlineto{\pgfqpoint{5.815000in}{0.926349in}}%
\pgfusepath{stroke}%
\end{pgfscope}%
\begin{pgfscope}%
\pgfpathrectangle{\pgfqpoint{0.842041in}{0.773252in}}{\pgfqpoint{4.972959in}{1.990267in}} %
\pgfusepath{clip}%
\pgfsetbuttcap%
\pgfsetroundjoin%
\pgfsetlinewidth{2.007500pt}%
\definecolor{currentstroke}{rgb}{0.400000,0.400000,0.400000}%
\pgfsetstrokecolor{currentstroke}%
\pgfsetdash{{7.400000pt}{3.200000pt}}{0.000000pt}%
\pgfpathmoveto{\pgfqpoint{2.386093in}{0.926349in}}%
\pgfpathlineto{\pgfqpoint{5.815000in}{0.926349in}}%
\pgfpathlineto{\pgfqpoint{5.815000in}{0.926349in}}%
\pgfusepath{stroke}%
\end{pgfscope}%
\begin{pgfscope}%
\pgfpathrectangle{\pgfqpoint{0.842041in}{0.773252in}}{\pgfqpoint{4.972959in}{1.990267in}} %
\pgfusepath{clip}%
\pgfsetbuttcap%
\pgfsetroundjoin%
\pgfsetlinewidth{1.003750pt}%
\definecolor{currentstroke}{rgb}{0.000000,0.000000,0.000000}%
\pgfsetstrokecolor{currentstroke}%
\pgfsetstrokeopacity{0.500000}%
\pgfsetdash{{3.700000pt}{1.600000pt}}{0.000000pt}%
\pgfpathmoveto{\pgfqpoint{2.386093in}{0.769918in}}%
\pgfpathlineto{\pgfqpoint{2.386093in}{2.766852in}}%
\pgfusepath{stroke}%
\end{pgfscope}%
\begin{pgfscope}%
\pgfpathrectangle{\pgfqpoint{0.842041in}{0.773252in}}{\pgfqpoint{4.972959in}{1.990267in}} %
\pgfusepath{clip}%
\pgfsetbuttcap%
\pgfsetroundjoin%
\pgfsetlinewidth{1.003750pt}%
\definecolor{currentstroke}{rgb}{0.000000,0.000000,0.000000}%
\pgfsetstrokecolor{currentstroke}%
\pgfsetstrokeopacity{0.500000}%
\pgfsetdash{{3.700000pt}{1.600000pt}}{0.000000pt}%
\pgfpathmoveto{\pgfqpoint{4.383769in}{0.769918in}}%
\pgfpathlineto{\pgfqpoint{4.383769in}{2.766852in}}%
\pgfusepath{stroke}%
\end{pgfscope}%
\begin{pgfscope}%
\pgfpathrectangle{\pgfqpoint{0.842041in}{0.773252in}}{\pgfqpoint{4.972959in}{1.990267in}} %
\pgfusepath{clip}%
\pgfsetrectcap%
\pgfsetroundjoin%
\pgfsetlinewidth{1.003750pt}%
\definecolor{currentstroke}{rgb}{0.000000,0.000000,0.000000}%
\pgfsetstrokecolor{currentstroke}%
\pgfsetdash{}{0pt}%
\pgfpathmoveto{\pgfqpoint{0.842041in}{0.773252in}}%
\pgfpathlineto{\pgfqpoint{0.842041in}{2.396085in}}%
\pgfusepath{stroke}%
\end{pgfscope}%
\begin{pgfscope}%
\pgfpathrectangle{\pgfqpoint{0.842041in}{0.773252in}}{\pgfqpoint{4.972959in}{1.990267in}} %
\pgfusepath{clip}%
\pgfsetbuttcap%
\pgfsetroundjoin%
\pgfsetlinewidth{1.003750pt}%
\definecolor{currentstroke}{rgb}{0.000000,0.000000,0.000000}%
\pgfsetstrokecolor{currentstroke}%
\pgfsetdash{{3.700000pt}{1.600000pt}}{0.000000pt}%
\pgfpathmoveto{\pgfqpoint{0.842041in}{2.396085in}}%
\pgfpathlineto{\pgfqpoint{0.842041in}{2.763519in}}%
\pgfusepath{stroke}%
\end{pgfscope}%
\begin{pgfscope}%
\pgfpathrectangle{\pgfqpoint{0.842041in}{0.773252in}}{\pgfqpoint{4.972959in}{1.990267in}} %
\pgfusepath{clip}%
\pgfsetrectcap%
\pgfsetroundjoin%
\pgfsetlinewidth{1.003750pt}%
\definecolor{currentstroke}{rgb}{0.000000,0.000000,0.000000}%
\pgfsetstrokecolor{currentstroke}%
\pgfsetdash{}{0pt}%
\pgfpathmoveto{\pgfqpoint{0.842041in}{2.763519in}}%
\pgfpathlineto{\pgfqpoint{0.842041in}{2.763519in}}%
\pgfusepath{stroke}%
\end{pgfscope}%
\begin{pgfscope}%
\pgfsetrectcap%
\pgfsetmiterjoin%
\pgfsetlinewidth{0.000000pt}%
\definecolor{currentstroke}{rgb}{0.000000,0.000000,0.000000}%
\pgfsetstrokecolor{currentstroke}%
\pgfsetstrokeopacity{0.000000}%
\pgfsetdash{}{0pt}%
\pgfpathmoveto{\pgfqpoint{0.842041in}{0.773252in}}%
\pgfpathlineto{\pgfqpoint{0.842041in}{2.763519in}}%
\pgfusepath{}%
\end{pgfscope}%
\begin{pgfscope}%
\pgfsetrectcap%
\pgfsetmiterjoin%
\pgfsetlinewidth{0.803000pt}%
\definecolor{currentstroke}{rgb}{0.000000,0.000000,0.000000}%
\pgfsetstrokecolor{currentstroke}%
\pgfsetdash{}{0pt}%
\pgfpathmoveto{\pgfqpoint{5.815000in}{0.773252in}}%
\pgfpathlineto{\pgfqpoint{5.815000in}{2.763519in}}%
\pgfusepath{stroke}%
\end{pgfscope}%
\begin{pgfscope}%
\pgfsetrectcap%
\pgfsetmiterjoin%
\pgfsetlinewidth{0.803000pt}%
\definecolor{currentstroke}{rgb}{0.000000,0.000000,0.000000}%
\pgfsetstrokecolor{currentstroke}%
\pgfsetdash{}{0pt}%
\pgfpathmoveto{\pgfqpoint{0.842041in}{0.773252in}}%
\pgfpathlineto{\pgfqpoint{5.815000in}{0.773252in}}%
\pgfusepath{stroke}%
\end{pgfscope}%
\begin{pgfscope}%
\pgfsetrectcap%
\pgfsetmiterjoin%
\pgfsetlinewidth{0.803000pt}%
\definecolor{currentstroke}{rgb}{0.000000,0.000000,0.000000}%
\pgfsetstrokecolor{currentstroke}%
\pgfsetdash{}{0pt}%
\pgfpathmoveto{\pgfqpoint{0.842041in}{2.763519in}}%
\pgfpathlineto{\pgfqpoint{5.815000in}{2.763519in}}%
\pgfusepath{stroke}%
\end{pgfscope}%
\begin{pgfscope}%
\pgftext[x=1.614067in,y=1.936792in,,base]{\rmfamily\fontsize{14.000000}{16.800000}\selectfont Stable}%
\end{pgfscope}%
\begin{pgfscope}%
\pgftext[x=3.384931in,y=1.936792in,,base]{\rmfamily\fontsize{14.000000}{16.800000}\selectfont Acquired stability}%
\end{pgfscope}%
\begin{pgfscope}%
\pgftext[x=5.099385in,y=1.936792in,,base]{\rmfamily\fontsize{14.000000}{16.800000}\selectfont Unstable}%
\end{pgfscope}%
\begin{pgfscope}%
\pgfsetbuttcap%
\pgfsetmiterjoin%
\definecolor{currentfill}{rgb}{1.000000,1.000000,1.000000}%
\pgfsetfillcolor{currentfill}%
\pgfsetfillopacity{0.600000}%
\pgfsetlinewidth{1.003750pt}%
\definecolor{currentstroke}{rgb}{1.000000,1.000000,1.000000}%
\pgfsetstrokecolor{currentstroke}%
\pgfsetstrokeopacity{0.600000}%
\pgfsetdash{}{0pt}%
\pgfpathmoveto{\pgfqpoint{3.445425in}{2.131656in}}%
\pgfpathlineto{\pgfqpoint{5.745556in}{2.131656in}}%
\pgfpathlineto{\pgfqpoint{5.745556in}{2.694074in}}%
\pgfpathlineto{\pgfqpoint{3.445425in}{2.694074in}}%
\pgfpathclose%
\pgfusepath{stroke,fill}%
\end{pgfscope}%
\begin{pgfscope}%
\pgfsetrectcap%
\pgfsetroundjoin%
\pgfsetlinewidth{2.007500pt}%
\definecolor{currentstroke}{rgb}{0.000000,0.000000,0.000000}%
\pgfsetstrokecolor{currentstroke}%
\pgfsetdash{}{0pt}%
\pgfpathmoveto{\pgfqpoint{3.500981in}{2.539398in}}%
\pgfpathlineto{\pgfqpoint{3.778759in}{2.539398in}}%
\pgfusepath{stroke}%
\end{pgfscope}%
\begin{pgfscope}%
\pgftext[x=3.889870in,y=2.490787in,left,base]{\rmfamily\fontsize{14.000000}{16.800000}\selectfont Long holding time}%
\end{pgfscope}%
\begin{pgfscope}%
\pgfsetbuttcap%
\pgfsetroundjoin%
\pgfsetlinewidth{2.007500pt}%
\definecolor{currentstroke}{rgb}{0.400000,0.400000,0.400000}%
\pgfsetstrokecolor{currentstroke}%
\pgfsetdash{{7.400000pt}{3.200000pt}}{0.000000pt}%
\pgfpathmoveto{\pgfqpoint{3.500981in}{2.279022in}}%
\pgfpathlineto{\pgfqpoint{3.778759in}{2.279022in}}%
\pgfusepath{stroke}%
\end{pgfscope}%
\begin{pgfscope}%
\pgftext[x=3.889870in,y=2.230411in,left,base]{\rmfamily\fontsize{14.000000}{16.800000}\selectfont No holding}%
\end{pgfscope}%
\end{pgfpicture}%
\makeatother%
\endgroup%